\documentclass[12pt]{article}
\usepackage{a4wide,epsfig,amsmath,amssymb,cite,scalefnt,graphicx}
\numberwithin{equation}{section}
\usepackage{slashed}
\usepackage{axodraw4j}
\usepackage{pstricks}
\usepackage{color}
\usepackage{changepage}
\usepackage{tabularx}
\usepackage{graphics}
\usepackage{array}

\def\npb{{\it{Nucl.\ Phys.}\ }{\bf B}}
\def\plb{{{\it Phys.\ Lett.}\ }{ \bf B}}

\def\be{\begin{equation}}
\def\ee{\end{equation}}
\def\bea{\begin{align}}
\def\eea{\end{align}}
\def\nn{\nonumber\\}

\def\tr{\hbox{tr}}

\def\tlambda{\tilde \lambda}
\def\ttheta{\tilde \theta}
\def\tTheta{\tilde \Theta}
\def\tLambda{\tilde \Lambda}
\def\yhat{\hat y}

\def\Gtil{\hat G}

\def\nutil{\tilde\nu}
\def\mutil{\tilde\mu}
\def\rhotil{\tilde\rho}
\def\sigtil{\tilde\sigma}

\def\Chat{\hat C}

\def\half{{\textstyle{{1}\over{2}}}}

\def\frakk[#1#2{{{#1}\over{#2}}}

\def\Atil{\tilde A}

\def\phibar{\bar\phi}

\def\atil{\tilde \alpha}

\def\dtil{\tilde \delta}

\def\gtil{\tilde \gamma}
\def\mtil{\tilde \mu}
\def\ntil{\tilde \nu}
\def\rtil{\tilde \rho}
\def\stil{\tilde \sigma}
\def\xtil{\tilde \xi}
\def\ztil{\tilde \zeta}

\def\Ncal{{\cal N}}

\def\pa{\partial}

\input amssym.def
\input amssym
\def\npb{Nucl. Phys. B}
\def\plb{Phys. Lett. B}
\def\pa{\partial}
\def\be{\begin{equation}}
\def\ee{\end{equation}}
\def\bea{\begin{align}}
\def\eea{\end{align}}
\def\nn{\nonumber\\}

\def\tr{\hbox{tr}}

\def \pr{\partial}
\def \d{{\rm d}}
\def \tr{{\rm tr }}

\def \bY{{\bar Y}}

\def \rO{{\rm O}}

\def \half{{\textstyle {1 \over 2}}}

\def \ts{\textstyle}
\def\del{{\rm d}}

\def \N{{\cal N}}

\def \d{{\rm d}}

\def \vphi{{\varphi}}

\def\btil{\tilde \beta}
\def\cirk{\,{\raise1pt \hbox{${\scriptscriptstyle \circ}$}}\,}

\def \olr{{\raise6.5pt\hbox{$\leftrightarrow  \! \! \! \! \!$}}}

\font\ninerm=cmr9 \font\ninesy=cmsy9
\font\eightrm=cmr8 \font\sixrm=cmr6
\font\eighti=cmmi8 \font\sixi=cmmi6
\font\eightsy=cmsy8 \font\sixsy=cmsy6
\font\eightbf=cmbx8 \font\sixbf=cmbx6
\font\eightit=cmti8
\def\eightpoint{\def\rm{\fam0\eightrm}
  \textfont0=\eightrm \scriptfont0=\sixrm \scriptscriptfont0=\fiverm
  \textfont1=\eighti  \scriptfont1=\sixi  \scriptscriptfont1=\fivei
  \textfont2=\eightsy \scriptfont2=\sixsy \scriptscriptfont2=\fivesy
  \textfont3=\tenex   \scriptfont3=\tenex \scriptscriptfont3=\tenex
  \textfont\itfam=\eightit  \def\it{\fam\itfam\eightit}%
  \textfont\bffam=\eightbf  \scriptfont\bffam=\sixbf
  \scriptscriptfont\bffam=\fivebf  \def\bf{\fam\bffam\eightbf}%
  \normalbaselineskip=9pt
  \setbox\strutbox=\hbox{\vrule height7pt depth2pt width0pt}%
  \let\big=\eightbig  \normalbaselines\rm}
\catcode`@=11 %
\def\eightbig#1{{\hbox{$\textfont0=\ninerm\textfont2=\ninesy
  \left#1\vbox to6.5pt{}\right.\n@@space$}}}
\def\vfootnote#1{\insert\footins\bgroup\eightpoint
  \interlinepenalty=\interfootnotelinepenalty
  \splittopskip=\ht\strutbox %
  \splitmaxdepth=\dp\strutbox %
  \leftskip=0pt \rightskip=0pt \spaceskip=0pt \xspaceskip=0pt
  \textindent{#1}\footstrut\futurelet\next\fo@t}
\catcode`@=12 %
\def\today{\number\day\ \ifcase\month\or January\or February\or March\or
April\or May\or June\or July\or
August\or September\or October\or November\or December\fi, \number\year}

\input epsf

\begin{document}

\begin{titlepage}
\begin{flushright}
LTH1017\\
\end{flushright}
\date{}
\vspace*{3mm}

\begin{center}
{\Huge The $a$-function for gauge theories}\\[12mm]
{\bf I.~Jack\footnote{{\tt dij@liv.ac.uk}} and
C.~Poole\footnote{{\tt c.poole@liv.ac.uk}}}\\

\vspace{5mm}
Dept. of Mathematical Sciences,
University of Liverpool, Liverpool L69 3BX, UK\\

\end{center}

\vspace{3mm}
\begin{abstract}
The $a$-function is a proposed quantity defined for quantum field theories
which
has a monotonic behaviour along renormalisation group flows, being related to
the $\beta$-functions via a gradient flow equation involving a positive
definite
metric.
We construct the $a$-function at four-loop order for a general gauge theory
with fermions and scalars, using only one and two loop $\beta$-functions;
we are then able to provide a stringent consistency check on the general
three-loop gauge $\beta$-function. In the case of an $\Ncal=1$ supersymmetric
gauge theory, we present a general condition on the chiral field anomalous
dimension which guarantees an exact all-orders expression for the $a$-function;
and we verify this up to fifth order (corresponding to the three-loop
anomalous dimension).
\end{abstract}

\vfill

\end{titlepage}


\section{Introduction}

It is natural to regard quantum field theories as points on a manifold with
the couplings $\{g^I\}$ as co-ordinates, and with a natural flow
determined by the $\beta$-functions $\beta^I(g)$. At fixed points the quantum
field theory is scale-invariant and is expected to become a conformal field
theory. It was suggested by Cardy\cite{Cardy}
that there might be a four-dimensional
generalisation of  Zamolodchikov's $c$-theorem\cite{Zam}
in two dimensions, such that
there is a function $a(g)$ which has monotonic behaviour under
renormalisation-group (RG) flow
(the strong $a$-theorem) or which
is defined at fixed points such that $a_{\rm UV}-a_{\rm IR}>0$ (the weak
$a$-theorem). It soon became clear that the coefficient (which we shall denote
$\frac14A$) of the Gauss-Bonnet
term in the trace of the energy-momentum tensor is the only natural candidate
for the $a$-function. A proof of the weak $a$-theorem has been presented by
Komargodski and Schwimmer\cite{KS} and further analysed and extended in
Refs.\cite{Luty,ElvangST}.

In other work, a perturbative version of the strong $a$-theorem has been
derived\cite{Analog}
from Wess-Zumino consistency conditions for the response of
the theory defined on curved spacetime, and with $x$-dependent couplings
$g^I(x)$, to a Weyl rescaling of the metric\cite{Weyl}.
This approach has been extended to other dimensions in
Refs.\cite{GrinsteinCKA,Nakthree}. The essential result is that
we can define a function $\Atil$ by
\be \Atil=A+W_I\beta^I \, ,
\ee
where $A$ is defined above and
$W_I$ is well-defined as an RG quantity on the theory extended as
described above, such that $\Atil$ satisfies the crucial equation
\be \pa_I\Atil =T_{IJ}\beta^J\, ,
\label{grad}
\ee
for
\be
T_{IJ} =G_{IJ}+2\pa_{[I}W_{J]}+2{\tilde \rho}_{[I}\cdot Q_{J]} \, .
\ee
Here $G_{IJ}=  G_{JI} $, ${\tilde \rho}_I$ and $Q_J$
 may all be computed perturbatively within the theory extended
to curved spacetime and $x$-dependent $g^I$;
for weak couplings $G_{IJ}$ can be shown to be positive definite in
four dimensions (in six dimensions, $G_{IJ}$ has recently been
computed to be negative definite at leading order\cite{Sternew}).
Eq.~(\ref{grad})
implies
\be
\mu \frac{d}{d\mu} \Atil=\beta^I\frac{\pa}{\pa g^I} \Atil=G_{IJ}
\beta^I\beta^J
\ee
thus verifying the strong $a$-theorem so long as $G_{IJ}$ is positive.
Crucially Eq.~(\ref{grad}) also imposes integrability conditions which
constrain the form of the $\beta$-functions and are the focus of this
paper. These conditions relate contributions to $\beta$-functions at different
loop orders.

We should
mention here that for theories with a global symmetry, $\beta^I$ in these
equations should be replaced by a $B^I$ which is defined, for instance,
in Ref.\cite{Analog}; however it was shown in Ref.\cite{Fortin, FortinC} that
the two quantities only begin to differ at three loops; and in
Ref.\cite{FortinS}
that there is no difference to all orders in supersymmetric theories
(see Ref.\cite{Nak} for related results). Hence
for our purposes in this article we may ignore the distinction.

The analysis
of Ref.\cite{Analog} was recently further extended in Ref.\cite{OsbJacnew}
(related results were also presented in Ref.\cite{Other}).
Expressions for the three-loop Yukawa $\beta$-functions and for the three-loop
contribution to the metric $G_{IJ}$ were derived for a general
fermion-scalar theory, and Eq.~(\ref{grad}) was checked up to this order;
indeed
it was shown that Eq.~(\ref{grad}) places strong constraints upon the form of
the $\beta$-function at this order. The $\Ncal=1$ supersymmetric Wess-Zumino
model was also considered as a special case. Moreover, here it was possible to
show that an exact formula for the $a$-function conjectured in
Refs.\cite{AnselmiAM, BarnesJJ, KutasovXU}
was valid at this order; in previous work\cite{SusyA} this had been checked
at the level of the two-loop $\beta$-functions for a general, gauged
$N=1$ supersymmetric theory (in the rest of this paper we shall describe this
as a ``two-loop check'' although the corresponding $a$-function is actually
a four-loop quantity). A sufficient
condition on the chiral field anomalous dimension for this exact $a$-function
to be viable was presented and shown (using the results
of Ref.\cite{pwb,Jack}) to be satisfied up to three loops for the
Wess-Zumino model.

Our goal in this article is to extend the work of Ref.\cite{OsbJacnew} to the
gauge
case. In the non-supersymmetric case, we show that the two-loop Yukawa
$\beta$-function for the most general renormalisable gauge theory coupled to
fermions and scalars is compatible with Eq.~(\ref{grad}); indeed by
imposing Eq.~(\ref{grad}) we are able to
obtain the terms in $\Atil$ containing scalar or Yukawa couplings
at this order up to only three free parameters, without any further
perturbative calculations--this approach has previously been applied in
Ref.\cite{Sann} to the case of the standard model. (Of course a gradient flow
for purely scalar quantum field theories was postulated some time ago by
Wallace and Zia\cite{Wallace}.)
As a bonus, we
show that this expression for $\Atil$ is also consistent with (and provides a
stringent check upon) the general {\it three}-loop gauge $\beta$-function
derived by Gracey, Jones and Pickering\cite{GJP}. As a further check,
we also compute $\Atil$ using the dimensional reduction version of the Yukawa
$\beta$-function and show that in the supersymmetric case,
this reduces to the version of $\Atil$
presented in Ref.\cite{SusyA} (and is hence compatible with
Ref.\cite{AnselmiAM,
BarnesJJ, KutasovXU}). Finally, again in the context of supersymmetry,
we extend the condition on the chiral superfield anomalous
dimension presented in Ref.~\cite{OsbJacnew} to the gauge case and show that
it is satisfied at three loops. We also derive the (very
non-trivial) constraints that this
condition imposes upon the form of the three-loop anomalous dimension.

\section{The non-supersymmetric case}
We consider a general renormalisable gauge theory with a simple
gauge group $G$  and $n_{\varphi}$
real scalars $\varphi_a$, $n_{\psi}$ two-component Weyl fermions $\psi_i$,
where $G\subset U(n_\psi) \cap O(n_\vphi)$.
For a Yukawa interaction
$\tfrac12\psi_i{\!}^T C  (Y_a)_{ij}\psi_j\,\varphi_a + \mbox{h.c.}$
and a quartic scalar interaction
$\tfrac{1}{4!}\lambda_{abcd}\varphi_a\varphi_b\varphi_c\varphi_d$ and with a
gauge coupling $g$ the basic couplings are then
\be
g^I\equiv\{g,Y_a,\bY_a, \lambda_{abcd}\}\, , \quad Y_a{}^T = Y_a \, , \quad
\bY_a = Y_a{\!}^* \, .
\ee
The (hermitian) gauge generators
for the scalar and fermion fields are denoted respectively
$t_A^{\vphi}= - t_A^{\vphi}{}^T$ and $t_A^{\psi}$,
$A=1\ldots n_V$, where $n_V = \dim G$,
and obey
\be
[t_A,t_B]=if_{ABC}t_C \, ,
\ee
and gauge invariance requires $Y_a \,t^\psi_A + t^\psi_A{}^T \, Y_a  =
t_A^\vphi{}_{ab} \, Y_b$, $t^\vphi_A{}_{ae} \lambda_{ebcd}
\vphi_a \vphi_b \vphi_c \vphi_d = 0$.
In order to simplify the form of our results,
it is convenient to assemble the Yukawa couplings
into a matrix
\be
y_a=\begin{pmatrix} Y_a&0\\0&\bY_a
\end{pmatrix},\qquad
\yhat_a=\begin{pmatrix} \bY_a&0\\0&Y_a
\end{pmatrix} = \sigma_1 y_a \sigma_1 \, ,
\label{ydef}
\ee
and
\be
T_A=\begin{pmatrix} t_A^{\psi}&0\\0&-t_A^{\psi*}
\end{pmatrix} \, , \qquad {\hat T}_A = \sigma_1 T_A \sigma_1 = - T_A{\!}^T \, .
\label{Tdef}
\ee
This corresponds to using the Majorana spinor $\Psi = \begin{pmatrix}
\psi_i  \\ - C^{-1} {\bar \psi}{}^i{}^T \end{pmatrix}$.

We should mention here that in our present calculations we have ignored
potential parity violating counterterms (i.e. containing
$\epsilon$-tensors). The analysis of Ref.\cite{Analog} was recently
extended\cite{KZ} to the case of theories with chiral anomalies, including the
possibility of parity violating anomalies. It would be interesting to carry out
the detailed computations necessary to exemplify the general conclusions of 
Ref.\cite{KZ}.

The one- and two-loop gauge $\beta$-functions are given by
\begin{align}
\beta_g&=-\beta_0g^3- \beta_1g^5 -
g^3 \,\frac{1}{2n_V}\tr[C^{\psi}\,\yhat_ay_a] \, ,\nn
\beta_0&=\tfrac13(11C_G-2R^{\psi}-\tfrac12R^{\vphi}) \, ,\nn
\beta_1&=\tfrac13C_G(34C_G-10R^{\psi}-R^{\vphi})
-\frac{1}{n_V}\tr[(C^{\psi})^2]-\frac{4}{n_V}\tr[(C^{\vphi})^2]\, ,
\label{betagone}
\end{align}
where
\begin{align}
\tr[t^{\psi}_At^{\psi}_B]= {}& R^{\psi}\delta_{AB},\qquad
\tr[t^{\vphi}_At^{\vphi}_B]=R^{\vphi}\delta_{AB},\nn
C^{\vphi}= {}& t^{\vphi}_At^{\vphi}_A, \qquad
C^{\psi}=T_AT_A \, , \quad
{\hat C}^{\psi}= C^\psi{\,}^T  \, ,
\label{Rdefs}
\end{align}
with $T_A$ as defined in Eq.~(\ref{Tdef}).
We follow Ref.~\cite{OsbJacnew} in removing factors of $1/16\pi^2$
which arise at each loop order by redefining
\be
\lambda_{abcd}\rightarrow 16\pi^2\lambda_{abcd} \, ,\quad
Y_a\rightarrow 4\pi Y_a,\quad g\rightarrow4\pi g\, .
\ee
The one-loop Yukawa $\beta$-function is given by
\be
\beta_{y\, a}^{(1)}
=2\, y_b\yhat_ay_b+\tfrac12[y_b\yhat_b-6g^2\Chat^{\psi}]y_a
+\tfrac12y_a[\yhat_by_b-6g^2C^{\psi}]+\tfrac12 \, \tr[y_a\yhat_b]y_b \, ,
\label{betayone}
\ee
and the one-loop scalar $\beta$-function is given by
\begin{align}
\tfrac{1}{4!} \, \beta^{(1)}_{\lambda\, abcd}\,
\vphi_a\vphi_b \vphi_c\vphi_d  = {}&
\big ( \tfrac18\, \lambda_{abef}\lambda_{cdef}+\tfrac32\, g^4
(t^{\vphi}_At^{\vphi}_B)_{ab}(t^{\vphi}_At^{\vphi}_B)_{cd}
-\tfrac12 \, \tr[y_a\yhat_by_c\yhat_d]\nn
&-\tfrac12\, \lambda_{abce}g^2(C^{\vphi})_{ed}
+\tfrac{1}{12}\, \lambda_{abce}\tr[y_e\yhat_d]\big )
\vphi_a\vphi_b \vphi_c\vphi_d \, .
\label{betasone}
\end{align}

The leading terms in the metric $G_{IJ}$ in Eq.~(\ref{grad}) may be written as
\cite{Analog}
\be \d s^2=G_{IJ}\del g^I\del g^J=2n_V\frac{1}{g^2}(1+\sigma g^2)(\d g)^2
+\tfrac{1}{6}\, \tr[\del\yhat_a\del y_a]
+\tfrac{1}{144}\, \del\lambda_{abcd}\del\lambda_{abcd} \, ,
\label{Tlow}
\ee
where $\sigma$ is given (using dimensional regularisation, DREG) by
\cite{Analog,SusyA}
\be
\sigma=\tfrac16\left(102C_G-20R^{\psi}-7R^{\vphi}\right) \, .
\label{Acalc}
\ee
We emphasise here that $y$ and $\yhat$ are not independent; and furthermore,
the result of a trace is unchanged by interchanging $y$ and $\yhat$.

The lowest-order contributions to $\Atil$
are given implicitly in Ref.~\cite{SusyA} as
\begin{align}
\Atil^{(2)}=&-n_V\beta_0g^2,\nn
\Atil^{(3)}=&-\tfrac12n_Vg^4(\beta_1+ \sigma \, \beta_0)
-\tfrac12g^2\tr[y_a\yhat_a\Chat^{\psi}]\nn
&+\tfrac{1}{24}\left(\tr[y_a\yhat_ay_b\yhat_b]
+2\, \tr[y_a\yhat_by_a\yhat_b]+\tfrac12\tr[y_a\yhat_b]\, \tr[y_a\yhat_b]\right).
\label{Alow}
\end{align}
They can easily be checked to satisfy Eq.~(\ref{grad}) with
Eqs.~(\ref{betagone}), (\ref{betayone}).

To proceed to the next order, we shall need the two-loop Yukawa
$\beta$-function
in addition to the one-loop scalar $\beta$-function in Eq.~(\ref{betasone}).
The two-loop $\beta$-function is given in general by
Refs.~\cite{Mach, JackO} in the form
\begin{align}
\beta_{y\,a}^{(2)} ={}& \sum_{\alpha=1}^{7}c_{\alpha}(G^y_{\alpha})_a
+\sum_{{\alpha}=8}^{19}c_{\alpha}[(G^y_{\alpha})_a+({\Gtil}^y_{\alpha})_a)\nn
&+c_{20}g^4(\Chat^{\psi}y_a+y_aC^{\psi})
+c_{21}g^4\Chat^{\psi}y_aC^{\psi}\nn
&+c_{22}g^4[(\Chat^{\psi})^2y_a+y_a(C^{\psi})^2]
+c_{23}g^4(C^{\vphi})_{ab}(\Chat^{\psi}y_b+y_bC^{\psi})\nn
&+\bigl(\tr[c_{24}\, y_a\yhat_b y_c\yhat_c+c_{25}\, y_a\yhat_cy_b \yhat_c
+c_{26}\, g^2\Chat^{\psi}y_a\yhat_b]\nn
& \qquad {} +c_{27}\, g^4(C^{\vphi2})_{ab}
+c_{28}g^4(C^{\vphi})_{ab}
+c_{29}\, \lambda_{acde}\lambda_{bcde}\bigr)y_b \, .
\label{betytwo}
\end{align}

\begin{table}[h]
\setlength{\extrarowheight}{1cm}
\setlength{\tabcolsep}{24pt}
\hspace*{-4cm}
\centering
\resizebox{6.7cm}{!}{
                \begin{tabular*}{20cm}{cccc}
                        \begin{picture}(162,133) (239,-159)
                        \SetWidth{1.0}
                        \SetColor{Black}
                        \Line(240,-123)(400,-123)
                        \Line[dash,dashsize=8.8](320,-123)(320,-27)
                        \Arc[dash,dashsize=9.6](304,-103)(52,-157.38,-22.62)
                        \Arc[dash,dashsize=8.2](312.38,-120.1)(24.552,-173.217,-76.767)
                        \Arc[dash,dashsize=7.4,clock](330.437,-146.78)(12.999,163.096,-59.676)
                        \Arc[dash,dashsize=9.2](342.376,-115.803)(42.424,-95.92,-11.141)
                        \end{picture}
                        &
                        \begin{picture}(162,151) (239,-159)
                        \SetWidth{1.0}
                        \SetColor{Black}
                        \Line(240,-105)(400,-105)
                        \Line[dash,dashsize=8.8](320,-105)(320,-9)
                        \Arc(320,-137)(20.616,129,489)
                        \Arc[dash,dashsize=8.6](314.875,-79.25)(63.758,-155.193,-104.418)
                        \Arc[dash,dashsize=8.6,clock](332.155,-84.723)(56.968,-21.931,-81.068)
                        \end{picture}
                        &
                        \begin{picture}(162,146) (239,-159)
                        \SetWidth{1.0}
                        \SetColor{Black}
                        \Line(240,-110)(400,-110)
                        \Line[dash,dashsize=8.8](320,-110)(320,-14)
                        \Arc[dash,dashsize=9](320.5,-90.667)(67.335,-163.314,-16.686)
                        \Arc[dash,dashsize=10](320.5,-101.333)(32.67,-164.617,-15.383)
                        \end{picture}
                        &
                        \begin{picture}(162,98) (239,-160)
                        \SetWidth{1.0}
                        \SetColor{Black}
                        \Line(240,-158)(400,-158)
                        \Line[dash,dashsize=8.8](320,-159)(320,-63)
                        \Arc[dash,dashsize=10.2,clock](320,-168.038)(49.038,168.188,11.812)
                        \end{picture}
                        \\
                        {\Huge $G^{y}_{1}$}
                        &
                        {\Huge $G^{y}_{2}$}
                        &
                        {\Huge $G^{y}_{3}$}
                        &
                        {\Huge $G^{y}_{4}$}
                        \\
                        \vspace*{1cm}
                        \\
                        \begin{picture}(162,134) (239,-161)
                        \SetWidth{1.0}
                        \SetColor{Black}
                        \Line(240,-122)(400,-122)
                        \Arc[dash,dashsize=10.2](320,-110.697)(49.313,-166.75,-13.25)
                        \EBox(309,-86)(331,-63)
                        \Line(309,-86)(331,-63)
                        \Line(331,-86)(309,-63)
                        \EBox(309,-86)(331,-63)
                        \Line[dash,dashsize=7.6](320,-122)(320,-86)
                        \Line[dash,dashsize=7.6](320,-63)(320,-28)
                        \end{picture}
                        &
                        \begin{picture}(162,141) (239,-160)
                        \SetWidth{1.0}
                        \SetColor{Black}
                        \Line(240,-115)(400,-115)
                        \Line[dash,dashsize=8.8](320,-116)(320,-20)
                        \EBox(310,-159)(332,-136)
                        \Line(310,-159)(332,-136)
                        \Line(332,-159)(310,-136)
                        \EBox(310,-159)(332,-136)
                        \Arc[dash,dashsize=8.8](313.375,-99.15)(50.905,-161.859,-92.674)
                        \Arc[dash,dashsize=8.8](330.696,-104.567)(46.49,-87.159,-12.968)
                        \end{picture}
                        &
                        \begin{picture}(162,113) (239,-144)
                        \SetWidth{1.0}
                        \SetColor{Black}
                        \Line(240,-143)(400,-143)
                        \EBox(309,-83)(331,-60)
                        \Line(309,-83)(331,-60)
                        \Line(331,-83)(309,-60)
                        \EBox(309,-83)(331,-60)
                        \Arc(320,-110)(16.279,137,497)
                        \Line[dash,dashsize=5](320,-143)(320,-126)
                        \Line[dash,dashsize=3.2](320,-94)(320,-83)
                        \Line[dash,dashsize=5.8](320,-60)(320,-32)
                        \end{picture}
                        &
                        \begin{picture}(162,98) (239,-159)
                        \SetWidth{1.0}
                        \SetColor{Black}
                        \Line(240,-158)(400,-158)
                        \Line[dash,dashsize=8.8](378,-158)(378,-62)
                        \Arc[dash,dashsize=10.4,clock](304,-168.038)(49.038,168.188,11.812)
                        \Arc[dash,dashsize=9,clock](305.5,-164.167)(24.296,165.296,14.704)
                        \end{picture}
                        \\
                        {\Huge $G^{y}_{5}$}
                        &
                        {\Huge $G^{y}_{6}$}
                        &
                        {\Huge $G^{y}_{7}$}
                        &
                        {\Huge $G^{y}_{8}$}
                        \\
                        \vspace*{1cm}
                        \\
                        \begin{picture}(162,127) (239,-159)
                        \SetWidth{1.0}
                        \SetColor{Black}
                        \Line(240,-129)(400,-129)
                        \Line[dash,dashsize=8.8](378,-129)(378,-33)
                        \Arc[dash,dashsize=9,clock](289,-133.478)(32.312,172.033,7.967)
                        \Arc[dash,dashsize=9](320.5,-127.1)(30.559,-176.435,-3.565)
                        \end{picture}
                        &
                        \begin{picture}(162,98) (239,-159)
                        \SetWidth{1.0}
                        \SetColor{Black}
                        \Line(240,-158)(400,-158)
                        \Line[dash,dashsize=8.8](378,-158)(378,-62)
                        \Arc(304,-117)(20.616,129,489)
                        \Arc[dash,dashsize=8.6,clock](292.863,-154.785)(40.99,-175.502,-256.076)
                        \Arc[dash,dashsize=8.4,clock](313.357,-157.071)(43.653,74.531,-1.219)
                        \end{picture}
                        &
                        \begin{picture}(162,145) (239,-160)
                        \SetWidth{1.0}
                        \SetColor{Black}
                        \Line(240,-111)(400,-111)
                        \Line[dash,dashsize=8.8](280,-112)(280,-16)
                        \Arc[dash,dashsize=10.2](320.5,-91.688)(67.329,-163.331,-16.669)
                        \Arc[dash,dashsize=10.2,clock](336.5,-110.5)(28.504,-178.995,-361.005)
                        \end{picture}
                        &
                        \reflectbox{\begin{picture}(162,108) (239,-159)
                        \SetWidth{1.0}
                        \SetColor{Black}
                        \Line(240,-148)(400,-148)
                        \Line[dash,dashsize=8.8](297,-148)(297,-52)
                        \Arc[dash,dashsize=10.2,clock](355.5,-146.5)(28.504,-178.995,-361.005)
                        \CBox(273.817,-157.817)(252.183,-136.183){Black}{Black}
                        \end{picture}}
                        \\
                        {\Huge $G^{y}_{9}$}
                        &
                        {\Huge $G^{y}_{10}$}
                        &
                        {\Huge $G^{y}_{11}$}
                        &
                        {\Huge $G^{y}_{12}$}
                        \\
                        \vspace*{1cm}
                        \\
                        \begin{picture}(162,108) (239,-159)
                        \SetWidth{1.0}
                        \SetColor{Black}
                        \Line(240,-148)(400,-148)
                        \Line[dash,dashsize=8.8](264,-148)(264,-52)
                        \Arc[dash,dashsize=10.2,clock](355.5,-146.5)(28.504,-178.995,-361.005)
                        \CBox(306.817,-157.817)(285.183,-136.183){Black}{Black}
                        \end{picture}
                        &
                        \reflectbox{\begin{picture}(162,108) (239,-159)
                        \SetWidth{1.0}
                        \SetColor{Black}
                        \Line(240,-148)(400,-148)
                        \Line[dash,dashsize=8.8](264,-148)(264,-52)
                        \Arc[dash,dashsize=9,clock](333,-159.75)(44.85,163.484,16.516)
                        \CBox(343.817,-157.817)(322.183,-136.183){Black}{Black}
                        \end{picture}}
                        &
                        \begin{picture}(162,107) (239,-160)
                        \SetWidth{1.0}
                        \SetColor{Black}
                        \Line[dash,dashsize=8.8](355,-149)(355,-54)
                        \EBox(283,-159)(305,-136)
                        \EBox(283,-159)(305,-136)
                        \Text(295,-147)[]{\huge{\Black{$A$}}}
                        \EBox(365,-159)(387,-136)
                        \EBox(365,-159)(387,-136)
                        \Text(377,-147)[]{\huge{\Black{$A$}}}
                        \Arc[dash,dashsize=10,clock](295.5,-150.525)(41.528,177.895,2.105)
                        \Line(240,-149)(283,-149)
                        \Line(305,-149)(365,-149)
                        \Line(387,-149)(400,-149)
                        \end{picture}
                        &
                        \begin{picture}(162,98) (239,-159)
                        \SetWidth{1.0}
                        \SetColor{Black}
                        \Line(240,-158)(400,-158)
                        \Line[dash,dashsize=8.8](368,-158)(368,-62)
                        \EBox(291,-134)(313,-111)
                        \Line(291,-134)(313,-111)
                        \Line(313,-134)(291,-111)
                        \EBox(291,-134)(313,-111)
                        \Arc[dash,dashsize=8,clock](302.209,-170.525)(50.778,165.72,102.753)
                        \Arc[dash,dashsize=8,clock](301.555,-168.149)(48.518,76.356,12.074)
                        \end{picture}
                        \\
                        {\Huge $G^{y}_{13}$}
                        &
                        {\Huge $G^{y}_{14}$}
                        &
                        {\Huge $G^{y}_{15}$}
                        &
                        {\Huge $G^{y}_{16}$}
                        \\
                        \vspace*{1cm}
                        \\
                        \begin{picture}(162,98) (239,-159)
                        \SetWidth{1.0}
                        \SetColor{Black}
                        \Line(240,-158)(400,-158)
                        \EBox(358,-120)(380,-97)
                        \Line(358,-120)(380,-97)
                        \Line(380,-120)(358,-97)
                        \EBox(358,-120)(380,-97)
                        \Line[dash,dashsize=7.6](369,-156)(369,-120)
                        \Line[dash,dashsize=7.6](369,-97)(369,-62)
                        \Arc[dash,dashsize=8.8,clock](297.5,-162.7)(39.779,173.214,6.786)
                        \end{picture}
                        &
                        \reflectbox{\begin{picture}(162,129) (239,-159)
                        \SetWidth{1.0}
                        \SetColor{Black}
                        \Line(240,-127)(400,-127)
                        \Line[dash,dashsize=8.8](332,-127)(332,-31)
                        \Arc[dash,dashsize=8.8](333,-113.25)(44.85,-163.484,-16.516)
                        \CBox(271.817,-136.817)(250.183,-115.183){Black}{Black}
                        \end{picture}}
                        &
                        \begin{picture}(162,132) (239,-159)
                        \SetWidth{1.0}
                        \SetColor{Black}
                        \Line(240,-124)(400,-124)
                        \Line[dash,dashsize=8.8](332,-124)(332,-28)
                        \Arc[dash,dashsize=9.4](318.5,-92.382)(65.62,-151.195,-28.805)
                        \CBox(308.817,-133.817)(287.183,-112.183){Black}{Black}
                        \end{picture}
                        &
                        \\
                        {\Huge $G^{y}_{17}$}
                        &
                        {\Huge $G^{y}_{18}$}
                        &
                        {\Huge $G^{y}_{19}$}
                        &
                \end{tabular*}
        }
\caption{The diagrams contributing to $\beta_{y\, a}^{(2)}$}
\label{fig1}
\end{table}

The contributions $G^y_{\alpha}$ are depicted in Table~(\ref{fig1});
$\Gtil^y_{\alpha}$ is the transpose of
$G^y_{\alpha}$. A solid or open box
represents $g^2 C^{\psi}$ or $g^2C^{\vphi}$ respectively. A box with a letter
``$A$'' represents the gauge generator $g T_A$.
Note that for each of $G^y_{\alpha}$, there is an alternation between
``hatted'' and ``unhatted'' $y$ matrices, as can be seen 
in Eq.~(\ref{betytwo}) for $G^y_{\alpha}$, $\alpha=20,\ldots 28$.
To give a couple of examples, $G^y_4$ represents
\be
(G_4^y)_a=\lambda_{abcd}\, y_b\yhat_cy_d \, ,
\ee
and $G^y_{19}$ corresponds to
\be
(G_{19}^y)_a=g^2y_b\, C^{\psi}\, \yhat_ay_b,\qquad
({\Gtil}^y_{19})_a=g^2y_b\yhat_a\, {\hat C}^{\psi}\,y_b \, .
\ee

We present here the results for the coefficients evaluated using standard
dimensional regularisation, DREG\cite{Mach, JackO}:
\begin{align}
c_1=2,\quad c_2=-1,\quad c_3&=-2,\quad c_{4}=-2, \quad c_5=-12,\nn
c_6=6,\quad
c_{7}=0,\quad
c_8=-\tfrac18,&\quad c_9=0,\quad c_{10}=-\tfrac{3}{8},\quad c_{11}=-1,
\nn c_{12}=0,\quad
c_{13}=-\tfrac74,\quad
c_{14}=-\tfrac{1}{4},\quad
c_{15}&=6,\quad
c_{16}=\tfrac92,\quad
c_{17}=0, \nn
c_{18}=3, \quad
c_{19}&=5,\quad
c_{20}=-\tfrac{1}{12}\left(194C_G-20R^{\psi}
-11R^{\vphi}\right),\nn
c_{21}=0,\quad c_{22}=-\tfrac32,\quad
c_{23}&=6,\quad c_{24}=-\tfrac34,\quad c_{25}=-\tfrac12,\nn
c_{26}=\tfrac52,\quad
c_{27}=-\tfrac{21}{2},&\quad
c_{28}=\tfrac{1}{12}(147C_G-12R^{\psi}-3R^{\vphi}),\quad
c_{29}=\tfrac{1}{12}.
\label{bcoeffs}
\end{align}
There are 33 coefficients altogether (counting $c_{20}$ and $c_{28}$ as
three each).
We do however have the freedom to redefine the couplings, corresponding to a
change in renormalisation scheme; at this order we may consider
\begin{align}
\delta y_a={} &\mu_1\, y_b\yhat_ay_b+\mu_2\, (y_b\yhat_by_a+y_a\yhat_by_b)
+\mu_3\, \tr[y_a\yhat_b]y_b \nn
&{}  +\mu_4\,g^2(\Chat^{\psi}y_a+y_aC^{\psi})
+\mu_5\, g^2C^{\phi}_{ab}y_b\, ,  \nn
\delta g={} &(\nu_1\, C_G+\nu_2\, R^{\psi}+\nu_3\, R^{\phi})g^3 \, .
\label{redefgen}
\end{align}
This results in a change in the $\beta$-function
\begin{align}
\delta\beta^{(2)}_{y\,a}=&
\left(\beta^{(1)}_y\cdot\frac{\pa}{\pa y}+
\beta^{(1)}_g\, \frac{\pa}{\pa g}\right)\delta y_a
-\left(\delta y\cdot\frac{\pa}{\pa y}
+\delta g\, \frac{\pa}{\pa g}\right)\beta^{(1)}_{y\,a} \, .
\label{delbet}
\end{align}
Using Eqs.~(\ref{betagone}), (\ref{betayone}) this leads to
\begin{align}
\delta c_2=\mu_1-4\mu_3, \quad
\delta c_6=-4\mu_5,&\quad
\delta c_7=-\mu_5,\quad
\delta c_9=-\mu_1+4\mu_2,\nn
\delta c_{10}=\mu_2-\mu_3,\quad
\delta c_{11}=\mu_1-4\mu_2, &\quad
\delta c_{13}=-6\mu_2-\mu_4,\quad
\delta c_{14}=-6\mu_2-\mu_4,\nn
\delta c_{16}=-\mu_5,\quad
\delta c_{19}=-6\mu_1-4\mu_4, &\quad
\delta c_{24}=-2\mu_2+2\mu_3,\quad
\delta c_{25}=-\mu_1+4\mu_3,\nn
\delta c_{26}={} &-12\mu_3-2\mu_4,\nn
\delta c_{20}={}&2C_G\left(3\nu_1-\tfrac{11}{3}\mu_4\right)
+2R^{\psi}\left(3\nu_2+\tfrac23\mu_4\right)
+R^{\phi}\left(6\nu_3+\tfrac13\mu_4\right),\nn
\delta c_{28}={}&-\tfrac13\left(22C_G-4R^{\psi}-R^{\phi}\right)\mu_5.
\label{redefres}
\end{align}
We observe that the redefinitions corresponding to $\mu_{1-4}$ are not all
independent; for instance we may remove $\mu_4$ by redefining
\begin{align}
\mu_1\rightarrow\mu_1-\tfrac23\mu_4,\quad
\mu_2&\rightarrow\mu_2-\tfrac16\mu_4,\quad
\mu_3\rightarrow\mu_3-\tfrac16\mu_4,\nn
\nu_1\rightarrow\nu_1+\tfrac{11}{9}\mu_4,\quad
\nu_2&\rightarrow\nu_2-\tfrac29\mu_4,\quad
\nu_3\rightarrow\nu_3-\tfrac{1}{18}\mu_4,
\end{align}
This is a general consequence of the form of the redefinition given by
Eq.~(\ref{delbet}), which implies that a redefinition
\be
\delta y_a=\beta^{(1)}_{y\,a}, \quad
\delta g=\beta^{(1)}_g
\ee
has no effect on $\beta^{(2)}_{y\,a}$;
however $\mu_5$ yields an independent redefinition
due to the fact that there happens to be no corresponding $C^{\phi}_{ab}y_b$
term in $\beta^{(1)}_{y\,a}$.
It then follows that $\mu_{1-5}$ and $\nu_{1-3}$ yield only 7
independent redefinitions; we therefore have $33-7=26$ independent
coefficients in the two-loop $\beta$-function.
Under the change Eq.
\eqref{redefgen}
\be
\delta \Atil^{(3)} = - \delta g \, \frac{\pa}{\pa g} \Atil^{(2)} \, ,
\ee
which corresponds to taking $\delta \sigma = 4
\left(\nu_1\, C_G+\nu_2\, R^{\psi}+\nu_3\, R^{\phi}\right)$
in Eq.~\eqref{Tlow}.

Applying Eq.~(\ref{grad}), we require $\Atil^{(4)}$ to satisfy
\begin{align}
\d_{y}\Atil^{(4)}={} &\d y \cdot T^{(3)}_{yy} \cdot \beta_y^{(1)}
+t_1\, g\tr[\d\yhat_ay_aC^{\psi}]\, \beta_g^{(1)}
+t_2\, g(C^{\vphi})_{ab}\tr[\d\yhat_ay_b ]\, \beta_g^{(1)}\nn
&+ \tfrac{1}{12}\, \tr[\d\yhat_a\beta_{y\, a}^{(2)}] \, ,\nn
\d_{\lambda}\Atil^{(4)}={} &\tfrac{1}{144}\, \d\lambda_{abcd}\,
\beta^{(1)}_{\lambda}{\!}_{abcd}   \, ,
\label{gradfour}
\end{align}

\begin{table}[h]
        \setlength{\extrarowheight}{1cm}
        \setlength{\tabcolsep}{24pt}
        \hspace*{-4cm}
        \centering
        \resizebox{6.7cm}{!}{
                \begin{tabular*}{20cm}{cccc}
                        \begin{picture}(162,171) (191,-159)
                        \SetWidth{1.0}
                        \SetColor{Black}
                        \Arc(272,-69)(80,143,503)
                        \Line(217,8)(240,-14)
                        \Line(217,-14)(240,8)
                        \Arc[dash,dashsize=9.6,clock](141.5,-68.5)(108.501,37.134,-37.134)
                        \Arc[dash,dashsize=9.6](410.873,-67.78)(114.879,145.675,215.2)
                        \CTri(204.373,-135)(227,-112.373)(249.627,-135){Black}{Black}\CTri(204.373,-135)(227,-157.627)(249.627,-135){Black}{Black}
                        \end{picture}
                        &
                        \begin{picture}(162,172) (191,-149)
                        \SetWidth{1.0}
                        \SetColor{Black}
                        \Arc(272,-68)(80,143,503)
                        \Line(217,9)(240,-13)
                        \Line(217,-13)(240,9)
                        \Arc[dash,dashsize=9.6,clock](141.5,-67.5)(108.501,37.134,-37.134)
                        \Arc[dash,dashsize=9.6](410.873,-66.78)(114.879,145.675,215.2)
                        \CTri(295.373,-1)(318,21.627)(340.627,-1){Black}{Black}\CTri(295.373,-1)(318,-23.627)(340.627,-1){Black}{Black}
                        \end{picture}
                        &
                        \begin{picture}(162,173) (191,-159)
                        \SetWidth{1.0}
                        \SetColor{Black}
                        \Arc(272,-67)(80,143,503)
                        \Line(217,10)(240,-12)
                        \Line(217,-12)(240,10)
                        \Arc[dash,dashsize=9.6,clock](141.5,-66.5)(108.501,37.134,-37.134)
                        \Arc[dash,dashsize=9.6](410.873,-65.78)(114.879,145.675,215.2)
                        \CTri(296.373,-135)(319,-112.373)(341.627,-135){Black}{Black}\CTri(296.373,-135)(319,-157.627)(341.627,-135){Black}{Black}
                        \end{picture}
                        &
                        \begin{picture}(162,194) (191,-147)
                        \SetWidth{1.0}
                        \SetColor{Black}
                        \Arc(272,-46)(80,143,503)
                        \Line(261,46)(284,24)
                        \Line(261,24)(284,46)
                        \Arc(272,-46)(35,143,503)
                        \Line[dash,dashsize=9.6](272,-11)(272,33)
                        \Line[dash,dashsize=9.6](272,-126)(272,-82)
                        \CTri(249.373,-123)(272,-100.373)(294.627,-123){Black}{Black}\CTri(249.373,-123)(272,-145.627)(294.627,-123){Black}{Black}
                        \end{picture}
                        \\
                        {\Huge $G^{T}_{1}$}
                        &
                        {\Huge $G^{T}_{2}$}
                        &
                        {\Huge $G^{T}_{3}$}
                        &
                        {\Huge $G^{T}_{4}$}
                        \\
                        \vspace*{1cm}
                        \\
                        \begin{picture}(162,174) (191,-147)
                        \SetWidth{1.0}
                        \SetColor{Black}
                        \Arc(272,-66)(80,143,503)
                        \Line(261,26)(284,4)
                        \Line(261,4)(284,26)
                        \Arc(272,-66)(35,143,503)
                        \Line[dash,dashsize=9.6](272,-31)(272,13)
                        \Line[dash,dashsize=9.6](272,-146)(272,-102)
                        \CTri(249.373,-93)(272,-70.373)(294.627,-93){Black}{Black}\CTri(249.373,-93)(272,-115.627)(294.627,-93){Black}{Black}
                        \end{picture}
                        &
                        \begin{picture}(162,174) (191,-147)
                        \SetWidth{1.0}
                        \SetColor{Black}
                        \Arc(272,-66)(80,143,503)
                        \Line(261,26)(284,4)
                        \Line(261,4)(284,26)
                        \Arc(272,-66)(35,143,503)
                        \Line[dash,dashsize=9.6](272,-31)(272,13)
                        \Line[dash,dashsize=9.6](272,-146)(272,-102)
                        \CTri(249.373,-39)(272,-16.373)(294.627,-39){Black}{Black}\CTri(249.373,-39)(272,-61.627)(294.627,-39){Black}{Black}
                        \end{picture}
                        &
                        \begin{picture}(162,207) (191,-147)
                        \SetWidth{1.0}
                        \SetColor{Black}
                        \Arc(272,-33)(80,143,503)
                        \Line(261,59)(284,37)
                        \Line(261,37)(284,59)
                        \Line[dash,dashsize=9.4](272,47)(272,-112)
                        \Arc[dash,dashsize=10](272,-99)(47,-180,0)
                        \CTri(249.373,-110)(272,-87.373)(294.627,-110){Black}{Black}\CTri(249.373,-110)(272,-132.627)(294.627,-110){Black}{Black}
                        \end{picture}
                        &
                        \begin{picture}(162,207) (191,-147)
                        \SetWidth{1.0}
                        \SetColor{Black}
                        \Arc(272,-33)(80,143,503)
                        \Line(261,59)(284,37)
                        \Line(261,37)(284,59)
                        \Line[dash,dashsize=9.4](272,47)(272,-112)
                        \Arc[dash,dashsize=10](272,-99)(47,-180,0)
                        \CTri(203.373,-98)(226,-75.373)(248.627,-98){Black}{Black}\CTri(203.373,-98)(226,-120.627)(248.627,-98){Black}{Black}
                        \end{picture}
                        \\
                        {\Huge $G^{T}_{5}$}
                        &
                        {\Huge $G^{T}_{6}$}
                        &
                        {\Huge $G^{T}_{7}$}
                        &
                        {\Huge $G^{T}_{8}$}
                        \\
                        \vspace*{1cm}
                        \\
                        &
                        \begin{picture}(172,194) (191,-147)
                        \SetWidth{1.0}
                        \SetColor{Black}
                        \Arc(272,-46)(80,143,503)
                        \SetWidth{0.0}
                        \CBox(362.314,-57.314)(339.686,-34.686){Black}{Black}
                        \SetWidth{1.0}
                        \Line(261,46)(284,24)
                        \Line(261,24)(284,46)
                        \Line[dash,dashsize=9.4](272,34)(272,-125)
                        \CTri(249.373,-123)(272,-100.373)(294.627,-123){Black}{Black}\CTri(249.373,-123)(272,-145.627)(294.627,-123){Black}{Black}
                        \end{picture}
                        &
                        \begin{picture}(162,195) (191,-147)
                        \SetWidth{1.0}
                        \SetColor{Black}
                        \Arc(272,-45)(80,143,503)
                        \Line(261,47)(284,25)
                        \Line(261,25)(284,47)
                        \EBox(262,-56)(284,-33)
                        \Line(262,-56)(284,-33)
                        \Line(284,-56)(262,-33)
                        \EBox(262,-56)(284,-33)
                        \Line[dash,dashsize=10](272,35)(272,-33)
                        \Line[dash,dashsize=10](272,-56)(272,-124)
                        \CTri(249.373,-123)(272,-100.373)(294.627,-123){Black}{Black}\CTri(249.373,-123)(272,-145.627)(294.627,-123){Black}{Black}
                        \end{picture}
                        &
                        \\
                        &
                        {\Huge $G^{T}_{9}$}
                        &
                        {\Huge $G^{T}_{10}$}
                        &
                \end{tabular*}
        }
        \caption{Contributions to $\d y \cdot T^{(3)}_{yy} \cdot \d ' y$}
        \label{fig2}
\end{table}

where $\d_y=\d y\cdot \frac{\pa}{\pa y}$, etc,
the lower-order metric contributions were read off from Eq.~({\ref{Tlow})
and we write
\be
\d y \cdot T^{(3)}_{yy} \cdot \d ' y
=\sum_{\alpha=1}^{10}T_{\alpha}G^T_{\alpha} \, ,
\label{TTdef}
\ee

The contributions to
$\d y \cdot T^{(3)}_{yy} \cdot \d ' y$ at this order
are depicted in Table~(\ref{fig2}). Here
a diamond represents $\d'y $ and a cross
$\d y$. As an example, $G^T_1$ represents
\be
G^T_1=\tr[\d\yhat_ay_b\yhat_b\d' y_a] \, .
\ee
$T^{(3)}_{yy}$ is symmetric up to the order at which we are working.
The $\beta$-functions $\beta_g^{(1)}$, $\beta^{(1)}_y$
in Eq.~(\ref{gradfour}) are given in
Eqs.~(\ref{betagone}), (\ref{betayone}).
There are no ``off-diagonal'' fermion-scalar
contributions to this order.
We parameterise $\Atil^{(4)}$ as
\be
\Atil^{(4)}=\sum_{\alpha=1}^{28}A_{\alpha}G^{A}_{\alpha} + {\rm O}(g^6) \, ,
\label{Afour}
\ee
where the different contributions $G^{A}_{\alpha}$ are depicted in
Table~(\ref{fig4}), with a similar notation to Table~(\ref{fig3}).
We have included $G_{28}^A$ as a reflection of the
general freedom to redefine
\be
\Atil\rightarrow \Atil+g_{IJ}\beta^I\beta^J \, ,
\label{ambig}
\ee

\begin{table}[h]
        \setlength{\extrarowheight}{1cm}
        \setlength{\tabcolsep}{24pt}
        \hspace*{-4cm}
        \centering
        \resizebox{5.8cm}{!}{
                \begin{tabular*}{20cm}{cccc}
                        \begin{picture}(162,162) (303,-207)
                        \SetWidth{1.0}
                        \SetColor{Black}
                        \Arc[dash,dashsize=10](377.677,-10.431)(101.765,-129.461,-43.903)
                        \Arc[dash,dashsize=10](545.132,-213.254)(161.295,126.143,177.422)
                        \Arc[dash,dashsize=10](225.213,-223.061)(159.701,6.133,57.082)
                        \Arc[dash,dashsize=10,clock](384.8,-126.4)(79.604,37.446,-90.576)
                        \Arc[dash,dashsize=10,clock](385.109,-125.284)(80.723,-90.787,-206.711)
                        \Arc[dash,dashsize=10,clock](383.587,-126.94)(81.019,152.077,33.689)
                        \end{picture}
                        &
                        \begin{picture}(162,162) (191,-159)
                        \SetWidth{1.0}
                        \SetColor{Black}
                        \Arc(272,-78)(80,143,503)
                        \Arc[dash,dashsize=10](272,-78)(43.6,127,487)
                        \Line[dash,dashsize=10](273,1)(273,-159)
                        \end{picture}
                        &
                        \begin{picture}(162,162) (191,-159)
                        \SetWidth{1.0}
                        \SetColor{Black}
                        \Arc(272,-78)(80,143,503)
                        \Line[dash,dashsize=10](273,1)(273,-159)
                        \Line[dash,dashsize=10](192,-79)(352,-79)
                        \end{picture}
                        &
                        \begin{picture}(162,162) (191,-159)
                        \SetWidth{1.0}
                        \SetColor{Black}
                        \Arc(272,-78)(80,143,503)
                        \Line[dash,dashsize=10](273,2)(272,-157)
                        \Arc[dash,dashsize=10](234.955,-31.596)(101.046,-103.131,0.338)
                        \Arc[dash,dashsize=10,clock](236.364,-163.283)(114.415,111.732,23.861)
                        \end{picture}
                        \\
                        {\Huge $G^{A}_{1}$}
                        &
                        {\Huge $G^{A}_{2}$}
                        &
                        {\Huge $G^{A}_{3}$}
                        &
                        {\Huge $G^{A}_{4}$}
                        \\
                        \vspace*{1cm}
                        \\
                        \begin{picture}(162,162) (191,-159)
                        \SetWidth{1.0}
                        \SetColor{Black}
                        \Arc(272,-78)(80,143,503)
                        \Arc[dash,dashsize=10,clock](271.5,-178.033)(78.035,142.009,37.991)
                        \Arc[dash,dashsize=10](203.358,-34.288)(60.603,-99.841,33.318)
                        \Arc[dash,dashsize=10](340.928,-38.442)(59.256,140.812,278.807)
                        \end{picture}
                        &
                        \begin{picture}(162,162) (191,-159)
                        \SetWidth{1.0}
                        \SetColor{Black}
                        \Arc(272,-78)(80,143,503)
                        \Line[dash,dashsize=9.4](272,1)(272,-157)
                        \Arc[dash,dashsize=9.6,clock](185.394,-79.246)(54.612,64.389,-63.22)
                        \Arc[dash,dashsize=9.6](357.661,-78)(52.661,114.289,245.711)
                        \end{picture}
                        &
                        \begin{picture}(162,162) (191,-159)
                        \SetWidth{1.0}
                        \SetColor{Black}
                        \Arc(272,-78)(80,143,503)
                        \Arc[dash,dashsize=10](202.047,-76.809)(52.954,-80.266,78.062)
                        \Arc(304,-79)(27.295,118,478)
                        \Line[dash,dashsize=10](304,-52)(304,-5)
                        \Line[dash,dashsize=10](304,-106)(305,-150)
                        \end{picture}
                        &
                        \begin{picture}(162,162) (191,-159)
                        \SetWidth{1.0}
                        \SetColor{Black}
                        \Arc(272,-78)(80,143,503)
                        \Arc(272,-44)(22.825,119,479)
                        \Arc(272,-113)(22.825,119,479)
                        \Line[dash,dashsize=4.8](272,-20)(272,2)
                        \Line[dash,dashsize=4.6](272,-68)(272,-89)
                        \Line[dash,dashsize=4.6](272,-158)(272,-136)
                        \end{picture}
                        \\
                        {\Huge $G^{A}_{5}$}
                        &
                        {\Huge $G^{A}_{6}$}
                        &
                        {\Huge $G^{A}_{7}$}
                        &
                        {\Huge $G^{A}_{8}$}
                        \\
                        \vspace*{1cm}
                        \\
                        \begin{picture}(171,172) (182,-159)
                        \SetWidth{1.0}
                        \SetColor{Black}
                        \Arc(272,-68)(80,143,503)
                        \Arc[dash,dashsize=10,clock](207.949,-67.5)(57.071,81.89,-81.89)
                        \Arc[dash,dashsize=9](214.194,-127.171)(31.131,119.215,329.437)
                        \Arc[dash,dashsize=10](339.804,-67.391)(57.826,102.793,258.221)
                        \end{picture}
                        &
                        \begin{picture}(168,188) (188,-159)
                        \SetWidth{1.0}
                        \SetColor{Black}
                        \Arc(272,-52)(80,143,503)
                        \Arc[dash,dashsize=10,clock](207.949,-51.5)(57.071,81.89,-81.89)
                        \Arc[dash,dashsize=9](272.145,-74.702)(83.298,164.185,376.531)
                        \Arc[dash,dashsize=10](339.804,-51.391)(57.826,102.793,258.221)
                        \end{picture}
                        &
                        \begin{picture}(162,162) (191,-159)
                        \SetWidth{1.0}
                        \SetColor{Black}
                        \Arc(272,-78)(80,143,503)
                        \Arc(272,-78)(29.698,135,495)
                        \Line[dash,dashsize=10](273,-47)(273,1)
                        \Line[dash,dashsize=10](272,-108)(272,-158)
                        \Line[dash,dashsize=8.6](243,-79)(302,-79)
                        \end{picture}
                        &
                        \begin{picture}(173,162) (180,-159)
                        \SetWidth{1.0}
                        \SetColor{Black}
                        \EBox(181,-90)(203,-67)
                        \Line(181,-90)(203,-67)
                        \Line(203,-90)(181,-67)
                        \EBox(181,-90)(203,-67)
                        \Arc[dash,dashsize=10](271.997,-78.003)(80.003,-89.282,89.998)
                        \Arc[dash,dashsize=10](185.274,-78)(118.726,-42.363,42.363)
                        \Arc[dash,dashsize=10,clock](357.25,-77.5)(117.251,-136.642,-223.358)
                        \Arc[dash,dashsize=10,clock](275.295,-85.168)(86.231,167.837,92.19)
                        \Arc[dash,dashsize=10](278.893,-68.832)(89.434,-166.309,-94.42)
                        \end{picture}
                        \\
                        {\Huge $G^{A}_{9}$}
                        &
                        {\Huge $G^{A}_{10}$}
                        &
                        {\Huge $G^{A}_{11}$}
                        &
                        {\Huge $G^{A}_{12}$}
                        \\
                        \vspace*{1cm}
                        \\
                        \begin{picture}(162,172) (191,-159)
                        \SetWidth{1.0}
                        \SetColor{Black}
                        \Arc(272,-68)(80,143,503)
                        \Arc[dash,dashsize=10](202.55,-33.983)(63.357,-95.934,41.502)
                        \Arc[dash,dashsize=10,clock](341.646,-36.392)(60.844,-84.952,-224.165)
                        \CBox(283.817,-157.817)(262.183,-136.183){Black}{Black}
                        \end{picture}
                        &
                        \begin{picture}(173,162) (191,-159)
                        \SetWidth{1.0}
                        \SetColor{Black}
                        \Arc(272,-78)(80,143,503)
                        \CBox(362.817,-87.817)(341.183,-66.183){Black}{Black}
                        \Arc[dash,dashsize=10](196.552,-77.466)(59.468,-71.927,68.859)
                        \Arc[dash,dashsize=10](347.364,-79.174)(60.364,108.711,250.284)
                        \end{picture}
                        &
                        \begin{picture}(162,162) (191,-159)
                        \SetWidth{1.0}
                        \SetColor{Black}
                        \Arc(272,-78)(80,143,503)
                        \EBox(286,-92)(308,-69)
                        \Line(286,-92)(308,-69)
                        \Line(308,-92)(286,-69)
                        \EBox(286,-92)(308,-69)
                        \Arc[dash,dashsize=8.8](343.27,-68.026)(46.281,107.958,178.729)
                        \Arc[dash,dashsize=8.8,clock](345.163,-84.617)(50.703,-107.4,-171.628)
                        \Arc[dash,dashsize=10,clock](194.417,-78.692)(61.597,69.489,-68.492)
                        \end{picture}
                        &
                        \begin{picture}(182,172) (182,-159)
                        \SetWidth{1.0}
                        \SetColor{Black}
                        \Arc(272,-68)(80,143,503)
                        \Arc[dash,dashsize=10,clock](207.949,-67.5)(57.071,81.89,-81.89)
                        \CBox(362.817,-78.817)(341.183,-57.183){Black}{Black}
                        \Arc[dash,dashsize=9](214.194,-127.171)(31.131,119.215,329.437)
                        \end{picture}
                        \\
                        {\Huge $G^{A}_{13}$}
                        &
                        {\Huge $G^{A}_{14}$}
                        &
                        {\Huge $G^{A}_{15}$}
                        &
                        {\Huge $G^{A}_{16}$}
                \end{tabular*}
        }
        \label{fig3}
\end{table}

\begin{table}[!h]
        \setlength{\extrarowheight}{1cm}
        \setlength{\tabcolsep}{24pt}
        \hspace*{-4cm}
        \centering
        \resizebox{5.8cm}{!}{
                \begin{tabular*}{20cm}{cccc}
                        \begin{picture}(162,195) (191,-159)
                        \SetWidth{1.0}
                        \SetColor{Black}
                        \Arc(272,-45)(80,143,503)
                        \EBox(261,-56)(283,-33)
                        \Line(261,-56)(283,-33)
                        \Line(283,-56)(261,-33)
                        \EBox(261,-56)(283,-33)
                        \Line[dash,dashsize=10](272,-33)(272,35)
                        \Line[dash,dashsize=10](272,-56)(272,-126)
                        \Arc[dash,dashsize=9.6](272,-120.805)(37.195,174.129,365.871)
                        \end{picture}
                        &
                        \begin{picture}(189,162) (177,-159)
                        \SetWidth{1.0}
                        \SetColor{Black}
                        \Arc[dash,dashsize=9.2,clock](204.5,-78.5)(55.502,79.095,-79.095)
                        \EBox(181,-89)(203,-66)
                        \EBox(181,-89)(203,-66)
                        \Text(193,-78)[]{\huge{\Black{$A$}}}
                        \EBox(340,-89)(362,-66)
                        \EBox(340,-90)(362,-67)
                        \Text(350,-78)[]{\huge{\Black{$A$}}}
                        \Arc[dash,dashsize=9](342.035,-78.5)(56.037,103.451,256.549)
                        \Arc[clock](271.5,-78.471)(80.472,171.085,8.915)
                        \Arc(272.072,-78.2)(79.806,-172.222,-8.503)
                        \end{picture}
                        &
                        \begin{picture}(170,162) (183,-159)
                        \SetWidth{1.0}
                        \SetColor{Black}
                        \Arc(272,-78)(80,143,503)
                        \CBox(205.817,-88.817)(184.183,-67.183){Black}{Black}
                        \Arc(271,-79)(33.015,125,485)
                        \Line[dash,dashsize=10](273,1)(271,-46)
                        \Line[dash,dashsize=10](272,-113)(272,-158)
                        \end{picture}
                        &
                        \begin{picture}(162,162) (191,-159)
                        \SetWidth{1.0}
                        \SetColor{Black}
                        \Arc(272,-78)(80,143,503)
                        \EBox(262,-57)(284,-34)
                        \Line(262,-57)(284,-34)
                        \Line(284,-57)(262,-34)
                        \EBox(262,-57)(284,-34)
                        \Arc(272,-105)(20.809,145,505)
                        \Line[dash,dashsize=7.8](273,-34)(273,1)
                        \Line[dash,dashsize=9.6](272,-57)(272,-84)
                        \Line[dash,dashsize=10.4](272,-126)(272,-158)
                        \end{picture}
                        \\
                        {\Huge $G^{A}_{17}$}
                        &
                        {\Huge $G^{A}_{18}$}
                        &
                        {\Huge $G^{A}_{19}$}
                        &
                        {\Huge $G^{A}_{20}$}
                        \\
                        \vspace*{1cm}
                        \\
                        \begin{picture}(163,135) (190,-221)
                        \SetWidth{1.0}
                        \SetColor{Black}
                        \EBox(229,-110)(251,-87)
                        \EBox(229,-110)(251,-87)
                        \EBox(288,-110)(310,-87)
                        \EBox(288,-110)(310,-87)
                        \EBox(229,-188)(251,-165)
                        \EBox(229,-188)(251,-165)
                        \EBox(288,-188)(310,-165)
                        \EBox(288,-188)(310,-165)
                        \Arc[dash,dashsize=6.6](232.658,-138.5)(41.661,95.037,264.963)
                        \Arc[dash,dashsize=7,clock](310.488,-138.488)(41.515,92.054,-90.674)
                        \Arc[dash,dashsize=4](187.232,-150.33)(81.861,9.372,38.833)
                        \Arc[dash,dashsize=4,clock](367.33,-159.84)(99.974,166.793,142.514)
                        \Arc[dash,dashsize=4,clock](191.739,-126.283)(78.001,-7.897,-40.558)
                        \Arc[dash,dashsize=4](325.041,-135.832)(56.053,-178.806,-132.739)
                        \Text(240,-99)[]{\huge{\Black{$B$}}}
                        \Text(240,-177)[]{\huge{\Black{$A$}}}
                        \Text(298,-177)[]{\huge{\Black{$A$}}}
                        \Text(299,-98)[]{\huge{\Black{$B$}}}
                        \end{picture}
                        &
                        \begin{picture}(172,162) (191,-159)
                        \SetWidth{1.0}
                        \SetColor{Black}
                        \Arc(272,-78)(80,143,503)
                        \CBox(361.817,-104.817)(340.183,-83.183){Black}{Black}
                        \Arc[dash,dashsize=10,clock](207.949,-77.5)(57.071,81.89,-81.89)
                        \CBox(361.817,-67.817)(340.183,-46.183){Black}{Black}
                        \end{picture}
                        &
                        \begin{picture}(183,162) (182,-159)
                        \SetWidth{1.0}
                        \SetColor{Black}
                        \Arc(272,-78)(80,143,503)
                        \CBox(204.817,-87.817)(183.183,-66.183){Black}{Black}
                        \CBox(363.817,-88.817)(342.183,-67.183){Black}{Black}
                        \Line[dash,dashsize=9.4](272,2)(272,-158)
                        \end{picture}
                        &
                        \begin{picture}(162,163) (191,-159)
                        \SetWidth{1.0}
                        \SetColor{Black}
                        \Arc(272,-77)(80,143,503)
                        \EBox(262,-61)(284,-38)
                        \Line(262,-61)(284,-38)
                        \Line(284,-61)(262,-38)
                        \EBox(262,-61)(284,-38)
                        \EBox(262,-116)(284,-93)
                        \Line(262,-116)(284,-93)
                        \Line(284,-116)(262,-93)
                        \EBox(262,-116)(284,-93)
                        \Line[dash,dashsize=8.4](272,2)(272,-38)
                        \Line[dash,dashsize=6.8](272,-61)(272,-93)
                        \Line[dash,dashsize=8.6](272,-116)(272,-158)
                        \end{picture}
                        \\
                        {\Huge $G^{A}_{21}$}
                        &
                        {\Huge $G^{A}_{22}$}
                        &
                        {\Huge $G^{A}_{23}$}
                        &
                        {\Huge $G^{A}_{24}$}
                        \\
                        \vspace*{1cm}
                        \\
                        \begin{picture}(172,163) (181,-159)
                        \SetWidth{1.0}
                        \SetColor{Black}
                        \Arc(272,-77)(80,143,503)
                        \CBox(203.817,-87.817)(182.183,-66.183){Black}{Black}
                        \EBox(261,-88)(283,-65)
                        \Line(261,-88)(283,-65)
                        \Line(283,-88)(261,-65)
                        \EBox(261,-88)(283,-65)
                        \Line[dash,dashsize=10](272,-65)(272,3)
                        \Line[dash,dashsize=10](272,-88)(272,-158)
                        \end{picture}
                        &
                        \begin{picture}(173,162) (191,-159)
                        \SetWidth{1.0}
                        \SetColor{Black}
                        \Arc(272,-78)(80,143,503)
                        \Arc[dash,dashsize=10,clock](207.949,-77.5)(57.071,81.89,-81.89)
                        \CBox(362.817,-88.817)(341.183,-67.183){Black}{Black}
                        \end{picture}
                        &
                        \begin{picture}(162,162) (191,-159)
                        \SetWidth{1.0}
                        \SetColor{Black}
                        \Arc(272,-78)(80,143,503)
                        \EBox(262,-90)(284,-67)
                        \Line(262,-90)(284,-67)
                        \Line(284,-90)(262,-67)
                        \EBox(262,-90)(284,-67)
                        \Line[dash,dashsize=9.6](272,1)(272,-67)
                        \Line[dash,dashsize=9.6](272,-90)(272,-158)
                        \end{picture}
                        &
                        \begin{picture}(162,203) (191,-138)
                        \SetWidth{1.0}
                        \SetColor{Black}
                        \Arc(272,-37)(80,143,503)
                        \Line[dash,dashsize=9.4](272,42)(272,-116)
                        \Text(289,45)[lb]{\huge{\Black{$(1) y$}}}
                        \Text(289,-143)[lb]{\huge{\Black{$(1) y$}}}
                        \CTri(249.373,41)(272,63.627)(294.627,41){Black}{Black}\CTri(249.373,41)(272,18.373)(294.627,41){Black}{Black}
                        \CTri(249.373,-114)(272,-91.373)(294.627,-114){Black}{Black}\CTri(249.373,-114)(272,-136.627)(294.627,-114){Black}{Black}
                        \end{picture}
                        \\
                        {\Huge $G^{A}_{25}$}
                        &
                        {\Huge $G^{A}_{26}$}
                        &
                        {\Huge $G^{A}_{27}$}
                        &
                        {\Huge $G^{A}_{28}$}
                \end{tabular*}
        }
        \caption{Contributions to $A^{(4)}$ in the non-supersymmetric case}
        \label{fig4}
\end{table}

\noindent together with a related redefinition of $T_{IJ}$; see
Ref.~\cite{OsbJacnew} for further details.
The purely $g$-dependent contributions to $\Atil^{(4)}$ of course cannot
be determined from Eq.~(\ref{gradfour}).
Eq.~(\ref{gradfour}) entails the system of equations

\begin{align}
6A_4=\tfrac16c_1,\quad 2A_{11}+4A_{28} = \tfrac12T_{8}+\tfrac{1}{6}c_2
=& \; 2T_{6}+2T_{5}+\tfrac{1}{6}c_{25}= 2T_{4}+\tfrac12T_{7},\nn
2A_{10}+8A_{28} = 2T_{7}+\tfrac{1}{6}c_3=T_8,&\quad 4A_3=\tfrac16c_4,\quad
2A_{17}= 2T_{10}+\tfrac{1}{6}c_5=\tfrac{1}{6}c_{6},\nn
4A_{20} = \tfrac12T_{10}+\tfrac{1}{6}c_{7},&\quad
4A_6+2A_{28} = \tfrac12(T_{2}+T_{3})+\tfrac13c_8=T_1,\nn
2A_{9}+8A_{28} = 2T_{2}+2T_{3}+\tfrac13c_9
=& \; T_{7}+\tfrac12T_{8}+\tfrac13c_{11}= 2T_{1}+\tfrac12T_{8},\nn
2A_7+2A_{28} = \tfrac12(T_{2}+T_{3})+\tfrac13c_{10}
=& \; \tfrac12T_{1}+T_{4} = T_{6}+T_{5}+\tfrac{1}{6}c_{24},\nn
2A_{14}-12A_{28} = -3T_{1}+\tfrac12T_{9}+\tfrac13c_{12}
=& -3T_{2}-3T_{3}+\tfrac13c_{14},\nn
4A_{13}-24A_{28} = -3(T_{1}+T_{2}+T_{3})
+\tfrac12T_{9}+\tfrac13c_{13}, &\quad
4A_{18}=\tfrac13c_{15},\quad
2A_{15} = \tfrac13c_{16}= T_{10}+\tfrac13c_{17},\nn
2A_{16}-48A_{28} = 2T_{9}-3T_{8}+\tfrac13c_{18}
=& -6T_{7}-3T_{8}+\tfrac13c_{19},\quad 2A_{26}=\tfrac13c_{20},\nn
36A_{28}+2A_{23} = -3T_{9}+\tfrac{1}{6}c_{21},\quad
36A_{28}+2A_{22} =& -3T_{9}+\tfrac{1}{3}c_{22},\quad
2A_{25} = -6T_{10}+\tfrac13c_{23}, \nn
2A_{19}-12A_{28} = -6T_{6}-6T_{5}+\tfrac{1}{6}c_{26}
=& -6T_{4}+\tfrac12T_{9},\nn 2A_{24}=\tfrac16c_{27},
\quad 2A_{27}=& \; \tfrac16c_{28},\quad 2A_2=\tfrac16c_{29},\nn
6A_{5}+3A_{28} =\tfrac12(T_{1}+T_{2}+T_{3}),
&\quad
6A_8+\tfrac32A_{28} = \tfrac12(T_{4}+T_{6}+T_{5})
\label{Aeqs}
\end{align}
where the $c_{\alpha}$ are given in Eq.~(\ref{bcoeffs}).
(The coefficients $A_1$,
$A_{12}$ and $A_{21}$ are determined immediately by
the second of Eqs.~\eqref{gradfour},
so we simply list their values later in Eqs.~(\ref{Afourres}).)
Solving Eqs.~(\ref{Aeqs}), we find the conditions
\begin{align}
T_2+T_3=2T_1-\tfrac23c_8,&\quad T_4=\tfrac12T_1-\tfrac{1}{3}c_8
+\tfrac13c_{10},\nn
T_5+T_6=T_1-\tfrac{1}{6}c_{24}-\tfrac{1}{3}c_8+\tfrac13c_{10},&\quad
T_7=2T_1-\tfrac13c_{11},\nn
T_{8}=4T_1-\tfrac83c_8+\tfrac23c_9,\quad
T_9=&-6T_1+2c_{24}+\tfrac13c_{26},\quad
T_{10}=\tfrac{1}{12}(c_6-c_5),
\label{Tcond}
\end{align}
 together with conditions on the $\beta$-function
coefficients
\begin{align}
c_2=-\tfrac14c_3-c_9+4c_{10},&\quad c_{11}=\tfrac14c_3+4c_8-c_9,\nn
 c_5-c_6=-4(c_{16}-c_{17}),&\quad c_{14}=\tfrac38c_3-\tfrac32c_9+c_{12}
-\tfrac14(c_{18}-c_{19})\nn
c_{24}=\tfrac{1}{8}c_3+2c_8-\tfrac12c_9+\tfrac12c_{25},&\quad
c_{26}=-\tfrac12(c_{18}-c_{19})-3c_{25}.
\label{ccond}
\end{align}
The conditions on $T_{1-6}$ in Eq.~(\ref{Tcond})
were already derived in Ref.~\cite{OsbJacnew}.
Reassuringly, the conditions Eq.~(\ref{ccond})
are satisfied by the coefficients in Eq.~(\ref{bcoeffs}), and also by
the redefinitions in Eq.~(\ref{redefres}). These six constraints in
principle leave only 19 of the 25 independent coefficients in the two-loop
$\beta$-function to be determined by perturbative computation.

It turns out that Eq.~(\ref{gradfour}) is sufficient to determine
the Yukawa or $\lambda$-dependent
part of $\Atil^{(4)}$ up to three free parameters;
here are the results for the case of dimensional regularisation:

\begin{align}
A_{1}=\tfrac{1}{144},\quad
A_{2}=\tfrac{1}{144},\quad
A_{3}=-\tfrac{1}{12},&\quad
A_{4}=\tfrac{1}{18},\quad
A_{5}=\tfrac{1}{144},\quad
A_6=A_{11}=0,\nn
A_7=-\tfrac{1}{24},\quad
A_8=-\tfrac{1}{288},\quad
A_{9}=\tfrac{1}{12},&\quad
A_{10}=\tfrac{1}{6},\quad
A_{12}=-\tfrac{1}{24},\quad
A_{13}=-\tfrac{7}{24},\nn
A_{14}=-\tfrac16,\quad
A_{15}=\tfrac34,\quad
A_{16}=-\tfrac23,&\quad
A_{17}=\tfrac12,\quad
A_{18}=\tfrac12,\quad
A_{19}=\tfrac{1}{12},\nn
A_{20}=\tfrac{3}{16},\quad
A_{21}=\tfrac14,\quad
A_{22}=\tfrac34,&\quad
A_{23}=1, \quad
A_{24}=-\tfrac{7}{8},\quad
A_{25}=-\tfrac72,\nn
A_{26}=&-\tfrac{1}{72}(194C_G-20R^{\psi}-11R^{\vphi})-t_1\beta_0,\nn
A_{27}=\tfrac{1}{144}(147C_G-12R^{\psi}-3R^{\vphi})-t_2\beta_0,&\quad A_{28}=\tfrac12 T_1=\alpha,
\label{Afourres}
\end{align}
where $\beta_0$ is given in Eq.~(\ref{betagone}). Since
$A_6$ only appears in Eq.~\eqref{Aeqs}
in the combination $4A_6+2A_{28}$, we have
set $A_6=0$ in line with Ref.~\cite{OsbJacnew}. We see that $A_{28}$ is arbitrary as expected, and we have parametrised this freedom by $\alpha$. We note that under
the redefinitions
in  Eq. \eqref{redefgen},
\be
\delta \Atil^{(4)} = - \left(\delta y\cdot\frac{\pa}{\pa y}
+\delta g\, \frac{\pa}{\pa g}\right) \Atil^{(3)}  + {\rm O}(g^6) \, .
\ee
Moreover the effect of these redefinitions on the metric coefficients in
Eq.~\eqref{gradfour} (as parametrised in Eq.~\eqref{TTdef})
may easily be computed using Eq.~\eqref{Tlow} as
\begin{align}
\delta T_1=\delta T_2=&\delta T_3=-\tfrac23\mu_2,\nn
\delta T_4=\delta T_5=&\delta T_6=-\tfrac13\mu_3,\nn
\delta T_7=&\tfrac12\delta T_8=-\tfrac13\mu_1,\nn
\delta T_9=-\tfrac23\mu_4,&\quad \delta T_{10}=-\tfrac13\mu_5,\nn
\delta t_1=-\tfrac43\mu_4,&\quad \delta t_2=-\tfrac23\mu_5.
\end{align}
Using Eq.~\eqref{redefres}), these results are easily seen
to agree with Eq.~\eqref{Tcond}.

It is remarkable that no knowledge of the ``metric'' coefficients
$T_{\alpha}$ is required to determine the $A_{\alpha}$ in this fashion;
of course the
$t_i$ in Eq.~(\ref{Afourres}), which define the ``off-diagonal''
fermion-gauge metric in Eq.~(\ref{gradfour}),
could be determined by a perturbative calculation if required, as was
accomplished for the fermion-scalar case in Ref.\cite{OsbJacnew}.
The results
in Eq.~(\ref{Afourres}) will be used in Sect. 3 in a check of the three-loop
$\beta_g$.

In Ref.\cite{OsbJacnew} the extension to three loops was accomplished by first
inferring the three-loop Yukawa $\beta$-function for a chiral fermion-scalar
theory, using the three-loop results derived in Ref.\cite{Chet} for the
standard
model, combined with the results for the supersymmetric Wess-Zumino model.
Such an approach will not work in the gauged case, unfortunately;
the
results of Ref.\cite{Chet} are only for the $SU(3)$ colour
gauge group, which of course
is not sufficient to determine how the three-loop Yukawa $\beta$-function
depends on a general gauge coupling.

\section{The three-loop gauge $\beta$-function}

The three-loop gauge $\beta$-function was computed in Ref.~\cite{GJP}
for a general gauge theory coupled to fermions and scalars.
In this section we shall show that our result for $\Atil^{(4)}$ is
compatible with this result via Eq.~(\ref{grad}). In fact, our result for
$\Atil^{(4)}$ determines the 16 terms in $\beta_g^{(3)}$ with Yukawa
couplings up to 4 (see later) unknown parameters. It is rather striking that
the two-loop calculation of $\beta_y^{(2)}$ (and the one-loop result
$\beta_{\lambda}^{(1)}$) have thereby provided
so much information on a three-loop RG quantity. This is an example of the
``$3-2-1$'' phenomenon noted in Refs.\cite{Sann, Sanna}; namely that
the gauge-gauge, fermion-fermion and scalar-scalar contributions to the
metric $G_{IJ}$ start at successive loop orders.

In our notation, $\beta_g^{(3)}$ can be written
\begin{align}
\tfrac{1}{g} \beta_g^{(3)}={} &\tfrac{1}{16n_V } \, \bigl(
-\tfrac{2}{3}G^A_{12}+G^A_{13}+3G^A_{14}+4G^A_{15}+12G^A_{16}-8G^A_{17}+8G^A_{18}\nn
&\qquad \  {} +7G^A_{19} -G^A_{20}+8G^A_{21}-10G^A_{22}
-2G^A_{23}\nn
&{} \qquad \ {} -28G^A_{24}-64G^A_{25}-48C_GG^A_{26}
+18C_GG^A_{27}+{\rm O}(g^6) \bigr) \, ,
\end{align}
where the $G^A_{\alpha}$ are implicitly defined in Table~(\ref{fig4}).
The purely $g$-dependent terms are not determined in this analysis.
It is then easy to show, using Eqs.~(\ref{Afour}), (\ref{Afourres}),
that we can write
\be
g\frac{\pa}{\pa g} \Atil^{(4)}=T^{(1)}_{gg}\beta_g^{(3)}
+T^{(2)}_{gg}\beta_g^{(2)}+T^{(3)}_{gg}\beta_g^{(1)}
+T^{(3)}_{gy}\cdot \beta_y^{(1)} \, ,
\label{betagthr}
\ee
in the form,
\begin{align}
g\frac{\pa}{\pa g} \Atil^{(4)}={}& 2n_V \, \tfrac{1}{g}\beta_g^{(3)}
+\tfrac13gn_V\left(102C_G-20R^{\psi}-7R^{\vphi}\right)\beta_g^{(2)}
+T^{(3)}_{gg}\beta_g^{(1)}\nn
&{} -g^2\tfrac{17}{12}\, \tr[\yhat_aC^{\vphi}\beta^{(1)}_{y\,a}]
+g^2(C^{\vphi})_{ab}\, \tr[\yhat_a\beta^{(1)}_{y\, b}]\, ,
\label{betagthra}
\end{align}
where $\beta_{g}^{(1)},\beta_g^{(2)}$ are given in Eq.~(\ref{betagone}).
We notice
that $T_{gg}^{(2)}$ agrees with the result for $\sigma$ in Eq.~(\ref{Acalc}).
$T_{gg}^{(3)}$ takes the form
\be
T_{gg}^{(3)}=
g\left(-\tfrac{10}{3}+4t_1\right)\tr[\yhat_a y_aC^{\psi}]
+g\left(-\tfrac12+4t_2\right)(C^{\vphi})_{ab}\tr[y_a\yhat_b]+{\rm O}(g^3) \, .
\label{betagthrb}
\ee
Unfortunately we have no means of disentangling the separate purely
$g$-dependent
contributions in $\Atil^{4}$ and in $T^{(3)}_{gg}\beta_g^{(1)}$,
without a three-loop calculation; but all the Yukawa or
$\lambda$ dependent contributions match exactly. If
\be
t_1=-\tfrac{17}{12}\, ,\qquad t_2=1 \, ,
\label{tsymm}
\ee
then we would have $T_{IJ}$ symmetric at this order; but as demonstrated in
Ref.\cite{OsbJacnew}, at three loops $T_{IJ}$ is not symmetric even for a
pure fermion-scalar theory for a general renormalisation scheme.

Had we not known $\beta_g^{(3)}$ then it would have been determined by
Eq.~\eqref{betagthr} up to the four parameters consisting of the
two coefficients in $T_{gy}^{(2)}$ and the two coefficients
in $T_{gg}^{(3)}$ (the values quoted for these
quantities in Eqs.~\eqref{betagthra}, \eqref{betagthrb}
of course deriving from our current knowledge of $\beta_g^{(3)}$).

\section{The supersymmetric case}

Here the analysis is extended to a general  $\N=1$
supersymmetric gauge theory, which may in principle be obtained from
the general non-supersymmetric theory discussed in Sect. 2 by an
appropriate choice of fields and couplings.
Such a theory can of course be rewritten in terms of $n_C$ chiral and
corresponding conjugate anti-chiral superfields, and indeed perturbative
computations are enormously simplified through the use of this formalism;
moreover, in the light of the non-renormalisation theorem and the
NSVZ formula\cite{nov,shifa}
for the exact gauge $\beta$-function, the renormalisation of the theory is
essentially entirely determined by the chiral superfield anomalous
dimension $\gamma$ (at least in a suitable renormalisation scheme).
In this section we shall therefore start anew using results
derived using superfield methods.
Nevertheless, in Sect 5 we show that (at least up two loops)
the results obtained using the two approaches match, as
indeed they must.

The crucial new feature in the supersymmetric context is the existence of a
proposed exact formula for the
$a$-function\cite{AnselmiAM, BarnesJJ, KutasovXU}. This exact form was
verified up to two loops in Ref.\cite{SusyA} for a general supersymmetric gauge
theory, and up to three loops\cite{OsbJacnew} in the case of the Wess-Zumino
model. Moreover in Ref.\cite{OsbJacnew} a sufficient condition on $\gamma$ to
guarantee the validity of this exact result was found and shown to be satisfied
up to three loops; related considerations appear in Refs.\cite{BarnesJJ,
KutasovXU},
see later for a discussion. In this section we shall generalise this condition
to the gauged case and check that it is satisfied up to three loops,
using the results of Ref.\cite{Jack}.

%

The couplings $g^I$ are now given by
$g^I =\{g,Y^{ijk},\bY_{ijk}\}$ with $\bY_{ijk} = (Y^{ijk})^*$.
The supersymmetric Yukawa $\beta$-functions
are expressible in terms of the anomalous dimension
matrix $\gamma_i{}^j$ in the form
\be
\beta_Y   =   Y * \gamma \, , \qquad
\beta_{\bY} = \gamma *  \bY \, ,
\ee
where for arbitrary $\omega_{\,i}{}^j$ we define
\begin{align}
(Y * \omega)^{ijk} \equiv  {}&
Y^{ljk}\omega_{\,l}{}^i + Y^{ilk}\omega_{\,l}{}^j +
Y^{ijl}\omega_{\,l}{}^k \, , \nn
(\omega * \bY)_{ijk}  \equiv {}&
\omega_i{}^l\, \bY_{ljk} +  \omega_j{}^l\, \bY_{ilk}
+  \omega_k{}^l \, \bY_{ijl} \, .
\end{align}
We also introduce a scalar product for Yukawa couplings\footnote{The
normalisation here is different from \cite{OsbJacnew}.}
\be
Y\cirk \bY=\bY \cirk Y = \tfrac16\, Y^{ijk}\bY_{ijk} \, ,
\ee
and it is further useful to define
\be
(\bY Y)_i{}^j = \tfrac12\, \bY_{ikl} Y{}^{jkl}  \quad \Rightarrow \quad
Y \cirk (\omega * \bY ) = (Y * \omega) \cirk \bY =   \tr\big ( ( \bY Y) \,
\omega \big )
 \, .
\label{lamdef}
\ee

The gauge $\beta$-function is assumed to have the form
\begin{align}
\beta_g =  f(g) \, \btil_g \, ,\qquad
\btil_g =  Q-2n_V^{-1}\, \tr[\gamma \, C_R] \, , \quad
f(g) = g^3 + {\rm O}(g^5) \,  ,
\label{betag}
\end{align}
where, with $R_A$ the gauge group generators,
\be
Q=T_R-3C_G,\quad
T_R\delta_{AB} = \tr(R_A R_B),\quad C_G\delta_{AB} = f_{ACD}f_{BCD},
\quad (C_R)_i{}^j = (R_A R_A)_i{}^j \, ,
\label{Ee}
\ee
and $n_V$ is the dimension of the gauge group.
For gauge invariance we must have
\be
Y * R_A = 0 \, , \qquad R_A *  \bY = 0 \, .
\ee

Under a change $g\to g'(g)= g + {\rm O}(g^3)$ then in Eq. \eqref{betag}
\be
f(g) \to f'(g') = \frac{\pr g'}{\pr g} \, f(g) \, , \qquad \gamma(g) \to
\gamma'(g')
= \gamma(g) \, ,
\label{varg}
\ee
assuming $g'$ is independent of $Y,\bY$. For an infinitesimal change
$ \delta f = f\,  \pr_g \delta g - \delta g \, \pr_g f $ and $ \delta \gamma = - \delta g \, \pr_g \gamma$.
The NSVZ form for the $\beta$-function is obtained if
\be
f(g) = \frac{g^3}{1-2\, C_G\, g^2} \, .
\ee
The resulting expression for $\beta_g$ originally appeared (for the special
case of
no chiral superfields) in Ref.~\cite{tim},
and was subsequently generalised, using instanton calculus,
in Ref.~\cite{nov}.
(See also Ref.~\cite{shifa}.) We note here that this result (called the
NSVZ form of $\beta_g$) is only valid in a specific renormalisation scheme,
which we correspondingly term the NSVZ scheme. The exact expression
generalises one and two-loop results obtained in Refs.~\cite{tja,pwa,tjlm}.
These results were computed using the dimensional reduction (DRED) scheme;
though in any case, the DRED and NSVZ schemes only part company at three
loops\cite{Schemedepa}.

The one and two-loop results for $\gamma$ are given
by\cite{tjlma,west}
\begin{align}
\gamma^{(1)} = {}& P \, ,\nn
\gamma^{(2)} = {}& -S_1-2g^2\, C_RP+2g^4\, Q\, C_R \, ,
\label{Eea}
\end{align}
where $P$ and $S_1$ are defined by
\begin{align}
P_i{}^j= {}& ( \bY Y )_i{}^j -2g^2(C_R)_i{}^j \, ,\nn
S_{1i}{}^j = {}& \bY_{ikn}Y^{jmn}P_m{}^k \, .
\label{Pdef}
\end{align}
We use here the notation and conventions of  Ref.~\cite{Jack}.


In the supersymmetric theory Eq.~(\ref{grad}) is assumed to now
take the form
\begin{align}
\del_Y {\tilde A}  = { }& \half \,  \del Y \cirk T_{\smash{Y\bY}}  \cirk
\beta_\bY
+\del Y \cirk T_{Yg} \, \btil_g  \, , \nn
\del_g{\tilde A}= {}& \del g\big(T_{gg}\, \btil_g + T_{gY} \cirk   \beta_Y
+ T_{\smash{g\bY}}\cirk \beta_\bY \big) \, ,
\label{ST}
\end{align}
(with a similar equation for $\del_{\bY}{\tilde A}$).
We have written the
RHS in terms of $\btil_g$, effectively absorbing the factor $f(g)$
in Eq. \eqref{betag} into $T_{Yg}$ and $T_{gg}$.
We omit potential $\beta_Y$ terms in the first of Eqs~(\ref{ST}) since
are not necessary to the order we shall consider.
For $\N=1$ supersymmetric theories there is, at critical points with
vanishing $\beta$-functions, an exact expression for $a$
\cite{AnselmiAM} in terms of the anomalous dimension matrix $\gamma$
or alternatively  the $R$-charge $R= \frac23(1+\gamma)$.
Introducing terms linear in $\beta$-functions there is
a corresponding expression which is valid  away from critical points
and this can then
be shown to  satisfy many of the properties associated with the
$a$-theorem \cite{BarnesJJ}, \cite{KutasovXU}.
For the theory considered here, with $n_C$ chiral scalar multiplets,
these results take the form
\be
{\tilde A} = {\ts{1\over 12}}(n_C+9\, n_V)  - \half \, \tr(\gamma^2) +
{\ts {1\over 3}} \, \tr(\gamma^3) + \Lambda \cirk \beta_\bY
+ n_V \, \lambda \, \btil_g +\beta_Y\cirk H \cirk \beta_{\bY}\, ,
\label{Aexact}
\ee
where $\btil_g$ is given by Eq.~(\ref{betag}) and we require
\be
\Lambda \cirk \beta_{\smash\bY} = \beta_Y \cirk {\bar \Lambda} \, .
\label{GaMB}
\ee
For the remainder of this section we omit for simplicity
the term involving $H$ in Eq.~(\ref{Aexact}); but return to it in Sect.~5.
In Refs.\cite{BarnesJJ} and
\cite{KutasovXU} $\Lambda,\lambda$ are Lagrange multipliers
enforcing constraints on the $R$-charges. At lowest order the
result for $\Lambda$ and also the metric $G$
obtained in Ref.\cite{BarnesJJ} are equivalent, up to matters
of definition and normalisation, with those obtained here. The
general form for $\Atil$ given by Eq.~(\ref{Aexact}) was 
verified up to two-loop order (for the
anomalous dimension) in Ref.~\cite{SusyA}.
$\Lambda$ may be constrained by imposing Eq.~(\ref{ST}).
Then
\begin{align}
\d_Y  \Atil
={} & \tr \big [  \d_Y \gamma  \; \big (
(\bY\Lambda )  -2\, \lambda \, C_R- \gamma + \gamma^2 \big ) \big  ] \nn
&{} +  ( \d_Y \Lambda )  \cirk \beta_\bY  + n_V \, \d_Y\lambda \; \btil_g\, .
\label{Dgam}
\end{align}
We also have
\begin{align}
\d_g  \Atil
 ={} & \tr \big [  \d_g \gamma  \; \big (
(\bY\Lambda )  -2\, \lambda \, C_R- \gamma + \gamma^2 \big ) \big  ] \nn
&+  ( \d_g \Lambda)   \cirk \beta_\bY  + n_V \,  \d_g \lambda \,  \btil_g  \, .
\label{Dg}
\end{align}
Hence if $\Lambda$, $\lambda$ are required to obey
\be
(\bY \Lambda ) - 2\, \lambda \, C_R
= \gamma - \gamma^2 + \Theta \cirk \beta_\bY  +\theta \, \btil_g\, ,
\label{detL}
\ee
where making  the indices explicit $\Theta \cirk \d \bY\to
\Theta_i{}^{j,klm} \d \bY_{klm}$ and $\theta \to \theta_i{}^j$,
Eq.~(\ref{Aexact})  then satisfies Eq.~(\ref{ST}) if we take
\begin{align}
\half \, \d Y \! \cirk T_{\smash{Y\bY}}   \cirk \d \bY& =
\tr \big [ \d_Y \gamma \; \Theta \cirk \d \bY \big ] +
 \d_Y \Lambda \cirk  \d \bY \, ,\nn
\del Y \cirk T_{Yg} &=\tr [ \del_Y\gamma \, \theta ]
+ n_V\, \del_Y\lambda \, ,\nn
\d g \, T_{gg}&=\tr\left[\d_g \gamma\, \theta\right]+
n_V \, \d_g \lambda \, , \nn
\d g \, T_{\smash{g\bY}} \cirk d \bY& =\ \tr\left[ \d_g \gamma\, \Theta\cirk
\del\bY\right] + \d_g \Lambda\cirk\d\bY \, .
\label{resGK}
\end{align}
Here $T_{gY}=0$. However from Eq. \eqref{GaMB}
\be
\d_g \Lambda \cirk \beta_\bY - \beta_Y \cirk \d_g {\bar \Lambda}
= \tr \big [ \d_g \gamma \big (  ({\bar \Lambda} Y) - (\bY \Lambda ) \big )
\big ] \, ,
\ee
which may be used to write Eq. \eqref{resGK} in equivalent forms with
non-zero $T_{gY}$.

A related result to Eq. \eqref{detL}, with effectively $\Theta,\theta=0$, is
contained in
Ref.\cite{BarnesJJ} and also discussed in Ref.\cite{KutasovXU}.
For supersymmetric theories,
satisfying Eq.~(\ref{detL}) is consequently essentially equivalent to
requiring Eq.~(\ref{ST}), although terms involving  $\Theta$
are necessary at higher orders. However, the work of
Refs.\cite{BarnesJJ, KutasovXU} implies that at least in the pure gauge case,
there may be renormalisation schemes in which $\theta$ may be set to zero.
It is striking that only minor modifications to the condition proposed
in Ref.~\cite{OsbJacnew} are required for the extension to the gauged case.

The condition (\ref{detL}) does not fully determine $\lambda, \theta$ since we
have the freedom
\begin{align}
\lambda \sim \lambda + \mu\,  \btil_g  \, , \qquad
\theta \sim  \theta - 2 \, \mu\,  C_R \, ,
\label{lam}
\end{align}
for arbitrary $\mu$. There is also a similar freedom in $\Lambda,\Theta$.

At lowest order $\Theta,\theta$ do not contribute
so that  (\ref{detL}) becomes
\be
(\bY\Lambda^{(1)})-2\, \lambda^{(1)} \, C_R=\gamma^{(1)} \, ,
\ee
and we may simply take from Eqs. \eqref{Eea},  \eqref{Pdef}
\be
\Lambda^{(1)}= Y \, , \qquad \lambda^{(1)} = g^2 \, ,
\label{lamonea}
\ee
from which
\be
\half \, \d Y \! \cirk T_{\smash{Y\bY}} {\!}^{(1)} \cirk \d \bY = \d Y \cirk \d
\bY \, , \qquad
T_{gg}{\!}^{(1)} = 2 \, n_V g\, .
\ee

At the next order we require
\be
(\bY\Lambda^{(2)})-2 \, \lambda^{(2)}C_R=\gamma^{(2)}-\gamma^{(1)2}
+\Theta^{(1)}\cirk \beta_\bY{\!}^{(1)} +\theta^{(1)} \, Q \, ,
\label{2order}
\ee
since $\btil_g{\!}^{(1)} = Q$, with $Q$ as defined in Eq.~(\ref{Ee}).
We may parameterise $\Lambda^{(2)}$ and
$\Theta^{(1)}$ by
\be
\Lambda^{(2)}=  \tLambda \, Y*P \, , \qquad
\Theta^{(1)} \cirk \d \bY =\tTheta \, (\d \bY \, Y) \, ,
\label{Lamthree}
\ee
since $  ( \bY \, Y*P ) =  S_1 + (\bY Y)\, P$ and
$(\beta_\bY{\!}^{(1)} Y) = (P* \bY  \, Y) =    S_1 + P \, (\bY Y)$
with $S_1$ as in Eq. \eqref{Pdef}.
We then find Eq.~(\ref{2order}) requires, since $(\bY Y) C_R = C_R (\bY Y)$,
\be
\tLambda - \tTheta = - 1 \, ,
\ee
and
\be
- 2 \, \lambda^{(2)}C_R - \theta^{(1)} \, Q = 2 \, g^4 \, Q C_R \, .
\ee
Hence
\be
\lambda^{(2)} =  \tlambda \, g^4 Q \, , \qquad
\theta^{(1)} = \ttheta  \, g^4 C_R \, ,
\ee
with
\be
2\, \tlambda + \ttheta = - 2 \, .
\label{lath}
\ee

\begin{table}[h]
        \setlength{\extrarowheight}{1cm}
        \setlength{\tabcolsep}{24pt}
        \hspace*{-4.5cm}
        \centering
        \resizebox{6.5cm}{!}{
                \begin{tabular*}{20cm}{cccc}
                        \begin{picture}(175,182) (191,-159)
                        \SetWidth{1.0}
                        \SetColor{Black}
                        \Arc(272,-58)(80,143,503)
                        \Line(272,21)(272,-137)
                        \SetWidth{0.0}
                        \Vertex(348,-42){16.971}
                        \Vertex(348,-88){16.971}
                        \SetWidth{1.0}
                        \CTri(249.373,-135)(272,-112.373)(294.627,-135){Black}{Black}\CTri(249.373,-135)(272,-157.627)(294.627,-135){Black}{Black}
                        \end{picture}
                        &
                        \begin{picture}(194,182) (175,-159)
                        \SetWidth{1.0}
                        \SetColor{Black}
                        \Arc(272,-58)(80,143,503)
                        \SetWidth{0.0}
                        \Vertex(351,-59){16.971}
                        \SetWidth{1.0}
                        \Line(272,21)(272,-137)
                        \SetWidth{0.0}
                        \Vertex(193,-59){16.971}
                        \SetWidth{1.0}
                        \CTri(249.373,-135)(272,-112.373)(294.627,-135){Black}{Black}\CTri(249.373,-135)(272,-157.627)(294.627,-135){Black}{Black}
                        \end{picture}
                        &
                        \begin{picture}(179,181) (191,-159)
                        \SetWidth{1.0}
                        \SetColor{Black}
                        \Arc(272,-59)(80,143,503)
                        \Line(272,20)(272,-138)
                        \SetWidth{0.0}
                        \Vertex(352,-59){16.971}
                        \SetWidth{1.0}
                        \Arc(381.561,-57.526)(78.563,133.987,227.018)
                        \CTri(249.373,-135)(272,-112.373)(294.627,-135){Black}{Black}\CTri(249.373,-135)(272,-157.627)(294.627,-135){Black}{Black}
                        \end{picture}
                        &
                        \begin{picture}(162,183) (191,-158)
                        \SetWidth{1.0}
                        \SetColor{Black}
                        \Arc(272,-57)(80,143,503)
                        \Arc[clock](220.79,-137.072)(111.636,102.304,66.11)
                        \Arc[clock](261.627,-92.294)(53.728,66.559,34.322)
                        \Arc[clock](285.384,-102.802)(40.039,40.123,-23.863)
                        \Arc(233.716,-14.692)(94.594,-105.146,9.549)
                        \Line(272,24)(272,-94)
                        \Line(272,-137)(272,-109)
                        \CTri(249.373,-134)(272,-111.373)(294.627,-134){Black}{Black}\CTri(249.373,-134)(272,-156.627)(294.627,-134){Black}{Black}
                        \end{picture}
                        \\
                        {\Huge $G^{\Lambda}_{1}$}
                        &
                        {\Huge $G^{\Lambda}_{2}$}
                        &
                        {\Huge $G^{\Lambda}_{3}$}
                        &
                        {\Huge $G^{\Lambda}_{4}$}
                        \\
                        \vspace*{1cm}
                        \\
                        \begin{picture}(175,182) (191,-159)
                        \SetWidth{1.0}
                        \SetColor{Black}
                        \Arc(272,-58)(80,143,503)
                        \Line(272,21)(272,-137)
                        \SetWidth{0.0}
                        \Vertex(348,-42){16.971}
                        \CBox(359.314,-100.314)(336.686,-77.686){Black}{Black}
                        \SetWidth{1.0}
                        \CTri(249.373,-135)(272,-112.373)(294.627,-135){Black}{Black}\CTri(249.373,-135)(272,-157.627)(294.627,-135){Black}{Black}
                        \end{picture}
                        &
                        \begin{picture}(188,181) (175,-159)
                        \SetWidth{1.0}
                        \SetColor{Black}
                        \Arc(272,-59)(80,143,503)
                        \Line(272,20)(272,-138)
                        \SetWidth{0.0}
                        \Vertex(193,-60){16.971}
                        \CBox(362.314,-72.314)(339.686,-49.686){Black}{Black}
                        \SetWidth{1.0}
                        \CTri(249.373,-135)(272,-112.373)(294.627,-135){Black}{Black}\CTri(249.373,-135)(272,-157.627)(294.627,-135){Black}{Black}
                        \end{picture}
                        &
                        \begin{picture}(173,182) (191,-159)
                        \SetWidth{1.0}
                        \SetColor{Black}
                        \Arc(272,-58)(80,143,503)
                        \Line(272,21)(272,-137)
                        \SetWidth{0.0}
                        \CBox(363.314,-69.314)(340.686,-46.686){Black}{Black}
                        \SetWidth{1.0}
                        \Arc(381.561,-56.526)(78.563,133.987,227.018)
                        \CTri(249.373,-135)(272,-112.373)(294.627,-135){Black}{Black}\CTri(249.373,-135)(272,-157.627)(294.627,-135){Black}{Black}
                        \end{picture}
                        &
                        \\
                        {\Huge $G^{\Lambda}_{5}$}
                        &
                        {\Huge $G^{\Lambda}_{6}$}
                        &
                        {\Huge $G^{\Lambda}_{7}$}
                        &
                        \\
                        \vspace*{1cm}
                        \\
                        \begin{picture}(169,182) (191,-159)
                        \SetWidth{1.0}
                        \SetColor{Black}
                        \Arc(272,-58)(80,143,503)
                        \Line(272,21)(272,-137)
                        \SetWidth{0.0}
                        \CBox(359.314,-100.314)(336.686,-77.686){Black}{Black}
                        \CBox(359.314,-54.314)(336.686,-31.686){Black}{Black}
                        \SetWidth{1.0}
                        \CTri(249.373,-135)(272,-112.373)(294.627,-135){Black}{Black}\CTri(249.373,-135)(272,-157.627)(294.627,-135){Black}{Black}
                        \end{picture}
                        &
                        \begin{picture}(182,182) (181,-159)
                        \SetWidth{1.0}
                        \SetColor{Black}
                        \Arc(272,-58)(80,143,503)
                        \Line(272,21)(272,-137)
                        \SetWidth{0.0}
                        \CBox(362.314,-71.314)(339.686,-48.686){Black}{Black}
                        \CBox(204.314,-71.314)(181.686,-48.686){Black}{Black}
                        \SetWidth{1.0}
                        \CTri(249.373,-135)(272,-112.373)(294.627,-135){Black}{Black}\CTri(249.373,-135)(272,-157.627)(294.627,-135){Black}{Black}
                        \end{picture}
                        &
                        &
                        \\
                        {\Huge $G^{\Lambda}_{8}$}
                        &
                        {\Huge $G^{\Lambda}_{9}$}
                        &
                        &
                \end{tabular*}
        }
        \caption{Contributions to $\Lambda^{(3)}\cirk\d \bY $}
        \label{fig5}
\end{table}

The freedom of choosing $\tlambda,\theta$ while satisfying Eq. \eqref{lath}
is a reflection of Eq. \eqref{lam}.
 From Eq.~(\ref{resGK})
\be
T_{gg}{\!}^{(2)}=4\tlambda \, n_V g^3 Q \, .
\ee
As a consequence of \eqref{lam} $\tlambda$ is arbitrary.
The computation in
Ref.\cite{SusyA} (specialising the DRED version
of Eq.~(\ref{Acalc}) to the supersymmetric case; and adjusting
for the differing definition of the ``$gg$''
metric) for $T_{gg}{\!}^{(2)}$ fixes
\be
\tlambda =-\tfrac52 \, ,
\label{lamone}
\ee
in this scheme.

At third order we require now in order to satisfy Eq.~(\ref{detL})
\begin{align}
(\bY\Lambda^{(3)})-2\, \lambda^{(3)}C_R ={}&
\gamma^{(3)}-\gamma^{(2)}\gamma^{(1)}
-\gamma^{(1)}\gamma^{(2)}
+\Theta^{(2)}\cirk \beta_\bY{\!}^{(1)} +\theta^{(2)} Q \nn
&{} +\Theta^{(1)}\cirk \beta_\bY{\!}^{(2)} +\theta^{(1)}\btil_g{\!}^{(2)},
\label{Lfoura}
\end{align}
where we write
\begin{align}
\Lambda^{(3)}\cirk\d \bY = {} &\sum_{\alpha=1}^{9}
\Lambda_{\alpha} G^{\Lambda}_{\alpha}\nn
&+ g^2(\Lambda_{10}Q+\Lambda_{11}C_G)\, (Y*P)\cirk \d \bY \,
+ g^4(\Lambda_{12}Q+\Lambda_{13}C_G)\, (Y*C_R)\cirk \d \bY\, ,
\end{align}
where $P$ is defined in Eq.~(\ref{Pdef}),
and the other distinct terms contributing to $\Lambda^{(3)}\cirk \d \bY $
are shown diagrammatically in Table~(\ref{fig5}). Here a ``blob''
represents
an insertion of the one-loop anomalous dimension.
The 3-point vertices alternate
between $Y$ and $\bY$. As an example, $G_6^{\Lambda}$ represents a
contribution
\be
\Lambda^{(3)}\cirk\del\bY
=g^2\, Y^{ikl}P_k{}^m (C_R)_l{}^n \d \bY_{imn} \, ,
\ee
and Eq.~(\ref{lamdef}) then
implies a contribution to $(\bY \Lambda^{(3)})$ of the form
\be
(\bY \Lambda^{(3)})_i{}^j= g^2 \big (   \bY_{imn}Y^{jkl}P_k{}^m(C_R)_l{}^n
+S_{1i}{}^k(C_R)_k{}^j +S_{2i}{}^kP_k{}^j\big ) \, .
\ee
Here $P,S_1$ are given in Eq.~(\ref{Pdef}) and  $S_2$ is defined by
\be
S_{2i}{}^j = \bY_{ikn}Y^{jmn} (C_R)_m{}^k \, .
\label{Sdef}
\ee
Similarly we write
\begin{align}
& \Theta^{(2)}\cirk \d \bY =\sum_{\alpha=1}^{6}\Theta_{\alpha}
G^{\Theta}_{\alpha}
+g^2\big (\Theta_{7}\, Q+\Theta_{8}\, C_G) \, (\d \bY Y \big ) \, ,\nn
& \lambda^{{(3)}}=g^4\tlambda_1\, \tr[PC_R]/n_V +g^6\left(
\tlambda_2 \, \tr [ C_R^2 ]/n_V + \tlambda_3\,Q^2
+\tlambda_4\,QC_G+\tlambda_5\, C_G^2 \right)\nn
&\theta^{(2)}=g^4 \ttheta_1\,S_2+g^4\ttheta_2\,PC_R
+g^6\big (\ttheta_3\, QC_R + \ttheta_4\, C_GC_R + \ttheta_5\, C_R^2\big ),
\end{align}
where the $G_\alpha^{\Theta}$
are shown diagrammatically in Table~(\ref{fig6}). A term in $S_1$ is
apparently possible in $\theta^{(2)}$ but is
excluded since there is no contribution to $\gamma^{(3)}$ involving
$g^2 Q S_1$. As a consequence of \eqref{lam} the resulting equations depend
only on $2\tlambda_3 + \ttheta_3, \, 2\tlambda_4 + \theta_4$.

\begin{table}[t]
\setlength{\extrarowheight}{1cm}
\setlength{\tabcolsep}{24pt}
\hspace*{-4cm}
\centering
\resizebox{8.5cm}{!}{
                \begin{tabular*}{25cm}{cccc}
                        \begin{picture}(226,98) (159,-191)
                        \SetWidth{0.0}
                        \SetColor{Black}
                        \Vertex(354,-142){16.971}
                        \SetWidth{1.0}
                        \Arc(272,-142)(47.802,143,503)
                        \Line(224,-142)(160,-142)
                        \Line(320,-142)(384,-142)
                        \CTri(207.373,-142)(230,-119.373)(252.627,-142){Black}{Black}\CTri(207.373,-142)(230,-164.627)(252.627,-142){Black}{Black}
                        \end{picture}
                        &
                        \begin{picture}(226,98) (159,-191)
                        \SetWidth{0.0}
                        \SetColor{Black}
                        \Vertex(185,-142){16.971}
                        \SetWidth{1.0}
                        \Arc(272,-142)(47.802,143,503)
                        \Line(224,-142)(160,-142)
                        \Line(320,-142)(384,-142)
                        \CTri(207.373,-142)(230,-119.373)(252.627,-142){Black}{Black}\CTri(207.373,-142)(230,-164.627)(252.627,-142){Black}{Black}
                        \end{picture}
                        &
                        \begin{picture}(226,113) (159,-176)
                        \SetWidth{0.0}
                        \SetColor{Black}
                        \Vertex(273,-81){16.971}
                        \SetWidth{1.0}
                        \Arc(272,-127)(47.802,143,503)
                        \Line(224,-127)(160,-127)
                        \Line(320,-127)(384,-127)
                        \CTri(207.373,-127)(230,-104.373)(252.627,-127){Black}{Black}\CTri(207.373,-127)(230,-149.627)(252.627,-127){Black}{Black}
                        \end{picture}
                        &
                        \begin{picture}(226,101) (159,-188)
                        \SetWidth{1.0}
                        \SetColor{Black}
                        \Arc(272,-139)(47.802,143,503)
                        \Line(224,-139)(160,-139)
                        \Line(320,-139)(384,-139)
                        \Arc(272,-70.367)(55.633,-133.082,-46.918)
                        \CTri(289.373,-111)(312,-88.373)(334.627,-111){Black}{Black}\CTri(289.373,-111)(312,-133.627)(334.627,-111){Black}{Black}
                        \end{picture}
                        \\
                        {\Huge $G^{\Theta}_{1}$}
                        &
                        {\Huge $G^{\Theta}_{2}$}
                        &
                        {\Huge $G^{\Theta}_{3}$}
                        &
                        {\Huge $G^{\Theta}_{4}$}
                        \\
                        \vspace*{1cm}
                        \\
                        &
                        \begin{picture}(226,98) (159,-191)
                        \SetWidth{1.0}
                        \SetColor{Black}
                        \Arc(272,-142)(47.802,143,503)
                        \Line(224,-142)(160,-142)
                        \Line(320,-142)(384,-142)
                        \SetWidth{0.0}
                        \CBox(364.314,-153.314)(341.686,-130.686){Black}{Black}
                        \SetWidth{1.0}
                        \CTri(207.373,-143)(230,-120.373)(252.627,-143){Black}{Black}\CTri(207.373,-143)(230,-165.627)(252.627,-143){Black}{Black}
                        \end{picture}
                        &
                        \begin{picture}(226,108) (159,-181)
                        \SetWidth{1.0}
                        \SetColor{Black}
                        \Arc(272,-132)(47.802,143,503)
                        \Line(224,-132)(160,-132)
                        \Line(320,-132)(384,-132)
                        \SetWidth{0.0}
                        \CBox(284.314,-96.314)(261.686,-73.686){Black}{Black}
                        \SetWidth{1.0}
                        \CTri(207.373,-132)(230,-109.373)(252.627,-132){Black}{Black}\CTri(207.373,-132)(230,-154.627)(252.627,-132){Black}{Black}
                        \end{picture}
                        &
                        \\
                        &
                        {\Huge $G^{\Theta}_{5}$}
                        &
                        {\Huge $G^{\Theta}_{6}$}
                        &
                \end{tabular*}
        }
\caption{Contributions to $\Theta^{(2)}\cirk \d \bY $}
\label{fig6}
\end{table}

We expand the three-loop anomalous dimension as
\be
\gamma^{(3)}=\sum_{\alpha=1}^{24}\gamma_{\alpha}G^{\gamma}_{\alpha},
\ee

\begin{table}[h]
        \setlength{\extrarowheight}{1cm}
        \setlength{\tabcolsep}{24pt}
        \hspace*{-4cm}
        \centering
        \resizebox{7.5cm}{!}{
                \begin{tabular*}{20cm}{cccc}
                        \begin{picture}(166,107) (189,-182)
                        \SetWidth{1.0}
                        \SetColor{Black}
                        \Arc(272,-133)(47.802,143,503)
                        \Line(224,-133)(190,-133)
                        \Line(320,-133)(354,-133)
                        \Vertex(252,-91){14.866}
                        \Vertex(295,-91){14.866}
                        \end{picture}
                        &
                        \begin{picture}(166,126) (189,-176)
                        \SetWidth{1.0}
                        \SetColor{Black}
                        \Arc(272,-114)(47.802,143,503)
                        \Line(224,-114)(190,-114)
                        \Line(320,-114)(354,-114)
                        \Vertex(272,-66){14.866}
                        \Vertex(272,-160){14.866}
                        \end{picture}
                        &
                        \begin{picture}(166,113) (189,-176)
                        \SetWidth{1.0}
                        \SetColor{Black}
                        \Arc(272,-127)(47.802,143,503)
                        \Line(224,-127)(190,-127)
                        \Line(320,-127)(354,-127)
                        \Vertex(272,-79){14.866}
                        \Arc(271.018,-61.29)(50.72,-135.246,-43.185)
                        \end{picture}
                        &
                        \begin{picture}(166,98) (189,-191)
                        \SetWidth{1.0}
                        \SetColor{Black}
                        \Arc(272,-142)(47.802,143,503)
                        \Line(224,-142)(190,-142)
                        \Line(320,-142)(354,-142)
                        \Line(238,-176)(301,-104)
                        \Line(239,-107)(265,-132)
                        \Line(278,-144)(310,-171)
                        \end{picture}
                        \\
                        {\Huge $G^{\gamma}_{1}$}
                        &
                        {\Huge $G^{\gamma}_{2}$}
                        &
                        {\Huge $G^{\gamma}_{3}$}
                        &
                        {\Huge $G^{\gamma}_{4}$}
                        \\
                        \vspace*{1cm}
                        \\
                        \begin{picture}(166,107) (189,-182)
                        \SetWidth{1.0}
                        \SetColor{Black}
                        \Arc(272,-133)(47.802,143,503)
                        \Line(224,-133)(190,-133)
                        \Line(320,-133)(354,-133)
                        \SetWidth{0.0}
                        \CBox(304.314,-101.314)(281.686,-78.686){Black}{Black}
                        \SetWidth{1.0}
                        \Vertex(252,-91){14.866}
                        \end{picture}
                        &
                        \begin{picture}(166,125) (189,-176)
                        \SetWidth{1.0}
                        \SetColor{Black}
                        \Arc(272,-115)(47.802,143,503)
                        \Line(224,-115)(190,-115)
                        \Line(320,-115)(354,-115)
                        \SetWidth{0.0}
                        \CBox(283.314,-175.314)(260.686,-152.686){Black}{Black}
                        \SetWidth{1.0}
                        \Vertex(272,-67){14.866}
                        \end{picture}
                        &
                        \begin{picture}(166,108) (189,-181)
                        \SetWidth{1.0}
                        \SetColor{Black}
                        \Arc(272,-132)(47.802,143,503)
                        \Line(224,-132)(190,-132)
                        \Line(320,-132)(354,-132)
                        \SetWidth{0.0}
                        \CBox(283.314,-96.314)(260.686,-73.686){Black}{Black}
                        \SetWidth{1.0}
                        \Arc(271.018,-66.29)(50.72,-135.246,-43.185)
                        \end{picture}
                        &
                        \\
                        {\Huge $G^{\gamma}_{5}$}
                        &
                        {\Huge $G^{\gamma}_{6}$}
                        &
                        {\Huge $G^{\gamma}_{7}$}
                        &
                        \\
                        \vspace*{1cm}
                        \\
                        \begin{picture}(166,104) (189,-185)
                        \SetWidth{1.0}
                        \SetColor{Black}
                        \Arc(272,-136)(47.802,143,503)
                        \Line(224,-136)(190,-136)
                        \Line(320,-136)(354,-136)
                        \SetWidth{0.0}
                        \CBox(304.314,-104.314)(281.686,-81.686){Black}{Black}
                        \CBox(265.314,-104.314)(242.686,-81.686){Black}{Black}
                        \end{picture}
                        &
                        \begin{picture}(166,120) (189,-181)
                        \SetWidth{1.0}
                        \SetColor{Black}
                        \Arc(272,-120)(47.802,143,503)
                        \Line(224,-120)(190,-120)
                        \Line(320,-120)(354,-120)
                        \SetWidth{0.0}
                        \CBox(283.314,-180.314)(260.686,-157.686){Black}{Black}
                        \CBox(283.314,-84.314)(260.686,-61.686){Black}{Black}
                        \end{picture}
                        &
                        &
                        \\
                        {\Huge $G^{\gamma}_{9}$}
                        &
                        {\Huge $G^{\gamma}_{10}$}
                        &
                        &
                \end{tabular*}
        }
        \caption{Contributions to $\gamma^{(3)}$}
        \label{fig7}
\end{table}

\noindent with
\begin{align}
G^{\gamma}_{8}=g^2S_1C_R,
\quad G^{\gamma}_{11}=g^4PC_R^2,&\quad
G^{\gamma}_{12}=g^4S_2C_R,\quad
G^{\gamma}_{13}=g^4C_GS_2,\nn
G^{\gamma}_{14}=g^4C_GPC_R,\quad
G^{\gamma}_{15}=g^4QPC_R,&\quad
G^{\gamma}_{16}=g^4QS_2,\quad
G^{\gamma}_{17}=g^4\tr[PC_R]/n_V \, C_R,\nn
G^{\gamma}_{18}=g^6C_R^3,\quad
G^{\gamma}_{19}=g^6C_GC_R^2,&\quad
G^{\gamma}_{20}=g^6QC_R^2, \quad
G^{\gamma}_{21}=g^6Q^2C_R,\nn
G^{\gamma}_{22}=g^6QC_GC_R,\quad
G^{\gamma}_{23}=&g^6C_G^2C_R,
\quad G^{\gamma}_{24}
=g^6\tr [ C_R^2 ]/n_V \, C_R,
\label{gamdef}
\end{align}
and with $Q$, $P$, $S_{1,2}$ as defined in Eqs.~(\ref{Ee}), (\ref{Pdef}),
(\ref{Sdef}).
The remainder of the distinct tensor contributions are depicted
in diagrammatic form in Table~(\ref{fig7}).
The basis for $\gamma^{(3)}$ is restricted by the absence of
one particle reducible contributions such as  $P^3$, $P^2C_R$, $S_{1,2}P$,
$PS_{1,2}$.

Using Eqs.~(\ref{Eea}), (\ref{Lamthree}) in Eq.~(\ref{Lfoura}) leads to a large
number
of consistency equations which constrain $\gamma^{(3)}$. If $g=0$ they reduce
to
\begin{align}
2\Lambda_{1} &= \gamma_{1}+\Theta_3+2\Theta_4= 2\Theta_{1}+2\Theta_2\, ,\nn
2\Lambda_2 &= 2\gamma_{2}+2\Theta_3
= 1+2\Theta_{1}-\tTheta \, ,\nn
2\Lambda_3 &= \gamma_{3}+2\Theta_4-\tTheta=
1+2\Theta_2+\Theta_3\, ,\nn
3\Lambda_{4} & =\gamma_4 \, ,
\label{Ysix}
\end{align}
which requires
\be
\gamma_{1}-2\gamma_{2}-\gamma_{3} = -2 \, .
\label{gzero}
\ee
These results were obtained in Ref. \cite{OsbJacnew}.
The other special case is for $Y,\bY=0$ when
\begin{align}
\gamma^{(3)} = {}& (\gamma_{18} - 2 \gamma_{11} )\, g^6  C_R^3
+\big (  ( \gamma_{19} - 2 \gamma_{14})\,  C_G
+  (\gamma_{20} - 2 \gamma_{15}) \, Q \big ) g^6  C_R^2 \nn
&{}+ \big ( \gamma_{21}\, Q^2 + \gamma_{22}\, QC_G + \gamma_{23}\, C_G^2
+ (\gamma_{24} - 2 \gamma_{17})\, \tr[C_R^2]/n_V  \big ) g^6 C_R\, .
\end{align}
In this case applying Eq. \eqref{Lfoura} with $\Lambda,\Theta\to 0$
it is necessary to require the conditions
\be
\gamma_{18} - 2 \gamma_{11} = -16 \, ,  \qquad  \gamma_{19} - 2 \gamma_{14} = 0
\, ,
\label{gsix}
\ee
as well as
\begin{align}
\gamma_{20} - 2 \gamma_{15} ={}&-8  -\ttheta_5 + 2 \ttheta_2 \, , \quad
\gamma_{21} = -2\tlambda_3 - \ttheta_3  \, ,\quad
\gamma_{22} = -2\tlambda_4 - \ttheta_4  \, , \nn
\gamma_{23} ={}&  -2\tlambda_5 \, , \quad
\gamma_{24}- 2\gamma_{17}   = 4 \tlambda_1 - 2 \tlambda_3    - 4 \ttheta \, .
\label{Yzero}
\end{align}
The relations in Eq. \eqref{gzero} were obtained in Refs. \cite{KutasovXU,
BarnesJJ}.

For the general case Eq. \eqref{Lfoura} implies additional relations
which further constrain $\gamma_\alpha$. From terms which start at
$\rO((Y\bY)^2)$
and using Eq. \eqref{Ysix}
\begin{align}
2\Lambda_5
&=  \gamma_{5}+\Theta_6+4\Theta_4-2\tTheta = 4+ 2\Theta_5-2\tTheta
\, ,\nn
\Lambda_{6} & = \gamma_{6}+\Theta_6= 0 \, ,\quad
2\Lambda_7 = \gamma_{7}= \Theta_6 \, ,\nn
\Lambda_{6}+4\Lambda_3& = \gamma_{8}+2\Theta_5
-2\tTheta \, , \nn
\Lambda_{10}& =\Theta_7  \, ,\quad
\Lambda_{11}=\Theta_8 \, ,
\label{Yfour}
\end{align}
which then entail, using Eq. \eqref{Ysix} to eliminate $\Lambda_3$,
\begin{align}
\gamma_8+ \gamma_{5}-\gamma_{6}
= 4 + 2 \gamma_3 \, ,\quad
\gamma_{6}+\gamma_{7}  = 0 \, .
\label{gtwo}
\end{align}
The remaining conditions arise from terms at $\rO(Y\bY)$ which become,
using Eq. \eqref{Yfour} to eliminate $\Lambda_5$,
\begin{align}
2\Lambda_8 & = \gamma_{9}= \gamma_{11} - 8 \, ,\nn
\Lambda_9 & = \gamma_{10}\, ,\quad
2\Lambda_9+4\Lambda_7 = \gamma_{12}\, ,\nn
\Lambda_{12}
&= \gamma_{16}+2\tTheta+\ttheta_1 \, , \nn
\Lambda_{12}+2\Lambda_{10}
 &= \gamma_{15} - 4 +2\Theta_7+2\tTheta +\ttheta_2\, ,\nn
\Lambda_{13} & =\gamma_{13} \, , \quad
\Lambda_{13}y+ 2\Lambda_{11}=\gamma_{14}+2\Theta_8 \, ,\nn
- 2 \tlambda_1 & = \gamma_{17} - 2 \ttheta \, ,
\label{Lamfour}
\end{align}
which then give
\begin{align}
2\gamma_{7}+2\gamma_{10} -\gamma_{12}=0 \,  ,\quad
\gamma_{9}-\gamma_{11}  =-8 \, ,  \quad
\gamma_{13}=\gamma_{14} \, .
\label{gfour}
\end{align}
As a consequence of Eq. \eqref{lam} the freedom $\delta \ttheta= - 2\mu$
requires also $\delta \tlambda_1 = - 2\mu$. We may combine Eq. \eqref{Yzero}
with Eq. \eqref{Lamfour} to give $\gamma_{24} = - 2 \tlambda_2$.

Altogether Eqs. \eqref{gzero}, \eqref{gsix}, \eqref{gtwo}, \eqref{gfour}
give eight conditions which
 are all satisfied by the coefficients as calculated\cite{pwb,Jack}:
\begin{align}
\gamma_{1}=-1,\quad
\gamma_{2}=-\tfrac12,\quad
\gamma_{3}&=2,\quad
\gamma_{4}=\tfrac14\kappa, \quad
\gamma_{5}=-2\kappa+4,\nn
\gamma_{6}=-\kappa, \quad
\gamma_{7}=\kappa,\quad
\gamma_{8}&=\kappa+4,\quad
\gamma_{9}=-5\kappa,\quad
\gamma_{10}=\kappa,\nn
\gamma_{11}=-5\kappa+8,\quad
\gamma_{12}=4\kappa,\quad
\gamma_{13}&=-\kappa,\quad
\gamma_{14}=-\kappa, \quad
\gamma_{15}=-2,\nn
\gamma_{16}=-4,\quad
\gamma_{17}=-12,\quad
\gamma_{18}&=-10\kappa,\quad
\gamma_{19}=-2\kappa,\quad
\gamma_{20}=-8,\nn
\gamma_{21}=2, \quad
\gamma_{22}=4\kappa+12, &\quad
\gamma_{23}=12\kappa,\quad
\gamma_{24}=-4\kappa,
\label{gamvals}
\end{align}
where $\kappa=6\zeta(3)$. As mentioned earlier, the NSVZ form
of the gauge $\beta$-function $\beta_g$ is valid only in a specific
renormalisation scheme (which differs from DRED at three loops). We are
therefore obliged for consistency to use the result for the anomalous
dimension corresponding to this NSVZ scheme. The required transformation was
presented in Ref.\cite{Schemedepa} and its effect on $\gamma^{(3)}$ given
in Ref.\cite{Schemedep}. In fact it is only $\gamma_{17}$
and $\gamma_{22}$ which are affected.

Once the conditions on the $\gamma_{\alpha}$ in
Eqs. \eqref{gzero}, \eqref{gsix}, \eqref{gtwo}, \eqref{gfour} are satisfied,
the $\Lambda_{\alpha}$, $\Theta_{\alpha}$ etc maybe be assigned in accord with
Eq.~(\ref{Lamfour}), with considerable arbitrariness; there is little to be
gained
from stating the residual relations amongst them.

In the Wess-Zumino case considered in  Ref.\cite{OsbJacnew}
the existence of an $a$-function satisfying Eq.~(\ref{grad}) implied that
$\gamma_{1}-2\gamma_{2}-\gamma_{3}$ was an invariant (in a sense described
in Ref.\cite{OsbJacnew}) but did not impose a specific value; thus
showing that Eq.~(\ref{detL}) is sufficient but not necessary. We might expect
similar remarks to apply to the other conditions in Eq.~(\ref{grad}). It is all
the
more striking that these conditions are in fact satisfied by the anomalous
dimension as computed.

We may count the independent parameters in the anomalous dimension as we
did in Section 2 for the Yukawa $\beta$-function. The essential Eqs.
\eqref{Aexact}
and \eqref{detL} are invariant under redefinitions of $g$ as in Eq.
\eqref{varg}
and taking $\delta Y = Y*h$ where
\be
 \delta \beta_Y = Y * \delta \gamma \, , \qquad
\delta \beta_\bY= \delta \gamma * \bY \, ,
\ee
and also assuming  $\delta \btil_g$ is given in terms of $\delta \gamma$  in
accord
Eq. \eqref{betag},  for
\be
\delta \gamma = - \big ( \delta g \partial_g + (Y*h) \cirk \pr_Y \big ) \gamma
+ \beta_\bY \cirk \pr_\bY h \, .
\ee
Taking
\begin{align}
h ={}&\atil \,S_1+\btil\, g^4S_2+\gtil \, g^2PC_R+\dtil \,g^4C_R^2
+(\ztil \, Q+\xtil \,C_G) g^4C_R \, , \nn
\delta g={} &g^5(\mtil \,Q^2+\ntil \,QC_G+\rtil \,C_G^2 +\stil \,
\tr[C_R^2]/n_V  )\, ,
\label{gamdefs}
\end{align}

\begin{table}[h]
        \setlength{\extrarowheight}{1cm}
        \setlength{\tabcolsep}{24pt}
        \hspace*{2cm}
        \centering
        \resizebox{11cm}{!}{
                \begin{tabular*}{20cm}{cc}
                        \begin{picture}(166,98) (189,-191)
                        \SetWidth{1.0}
                        \SetColor{Black}
                        \Arc(272,-142)(47.802,143,503)
                        \Line(224,-142)(190,-142)
                        \Line(320,-142)(354,-142)
                        \Photon(239,-108)(309,-172){5}{4.5}
                        \Photon(239,-176)(267,-147){5}{2.5}
                        \Photon(281,-132)(306,-109){5}{2}
                        \end{picture}
                        &
                        \begin{picture}(166,106) (189,-183)
                        \SetWidth{1.0}
                        \SetColor{Black}
                        \Arc(249,-134)(47.802,143,503)
                        \Line(201,-134)(190,-134)
                        \Line(297,-134)(354,-134)
                        \PhotonArc[clock](274.259,-121.307)(38.871,114.727,-19.058){5}{4.5}
                        \Photon(285,-103)(298,-101){5}{1}
                        \PhotonArc[clock](312.298,-124.954)(29.141,96.499,-18.084){5}{3.5}
                        \end{picture}
                        \\
                        {\LARGE $G^{NP}_{1}$}
                        &
                        {\LARGE $G^{NP}_{2}$}
        \end{tabular*}
        }
\caption{Non planar Feynman diagrams used to define $\gamma^{\prime(3)}$}
\label{fig9}
\end{table}

results in
\begin{align}
\delta\gamma_1=2\atil,\quad
\delta\gamma_2=\atil,&\quad
\delta\gamma_5=2\atil+\btil-\gtil,\quad
\delta\gamma_6=\btil,\nn
\delta\gamma_7=-\btil,\quad
\delta\gamma_8=\gtil-2\atil,&\quad
\delta\gamma_{9}=-\dtil,\quad
\delta\gamma_{11}=-\dtil,\nn
\delta\gamma_{12}=-2\btil,\quad
\delta\gamma_{13}=-\xtil,&\quad
\delta\gamma_{14}=-\xtil,\quad
\delta\gamma_{15}=-\ztil,\nn
\delta\gamma_{16}=-\ztil,\quad
\delta\gamma_{18}=-2\dtil,&\quad
\delta\gamma_{19}=-2\xtil,\quad
\delta\gamma_{20}=-2\ztil,\nn
\delta\gamma_{21}=4\mutil,\quad
\delta\gamma_{22}=4\nutil,&\quad
\delta\gamma_{23}=4\rhotil,\quad
\delta\gamma_{24}=4\sigtil ,
\end{align}
which leave Eqs. \eqref{gzero}, \eqref{gsix}, \eqref{gtwo}, \eqref{gfour}
invariant.
Allowing for such variations the
number of independent parameters is therefore $24-10=14$ but the eight
constraints in Eqs.~(\ref{gzero}), (\ref{gsix}),  (\ref{gtwo}), (\ref{gfour})
reduce
the number of free parameters in $\gamma^{(3)}$ to be reduced to six.

In the $g=0$ case, the only coefficient in $\gamma^{(3)}$ with a
$\kappa$-dependence, $\gamma_4$, corresponds to a non-planar graph. In
the general case there is no such obvious association between
non-planar Feynman graphs and coefficients in $\gamma^{(3)}$ with
$\kappa$-dependence (evaluated using DRED).
However, an  intriguing observation is that a redefinition given by choosing
\begin{align}
\btil=\kappa,\quad \gtil=&-\kappa, \quad \dtil=-2\kappa, \nn
\nutil=-\kappa,\quad \rhotil=&-3\kappa,\quad \sigtil=\kappa \, ,
\end{align}
(and the remaining coefficients in Eq.~\eqref{gamdefs} set to zero)
gives a redefined $\gamma^{(3)}$
\be
\gamma^{\prime (3)}=\gamma^{(3)}|_{\kappa=0}
+\gamma_4G^{\gamma}_4+\kappa G^{\rm{NP}}_1+2\kappa G^{\rm{NP}}_2 \, ,
\ee
where
\begin{align}
G^{\rm{NP}}_1= {}&G^{\gamma}_9+G^{\gamma}_{10}+g^4(-2C_RS_2+PC_R^2
+C_GS_2-C_GPC_R)+2g^6(C_R^3-C_GC_R^2),\nn
G^{\rm{NP}}_2= {}&G^{\gamma}_9-G^{\gamma}_{10}+g^4(PC_R^2+C_GPC_R)
+2g^6(C_R^3+C_GC_R^2) \, ,
\end{align}
are the contributions corresponding to the Feynman diagrams
shown in Table~(\ref{fig9}).
The implication is that there is a scheme in which the
$\kappa$-dependent terms in $\gamma^{(3)}$ are generated solely by
non-planar diagrams.

\section{Reduction of non-supersymmetric results to supersymmetric case}

In this section we shall check that the $a$-function obtained using the methods
of Section 2 for a general theory is compatible, upon specialisation to the
supersymmetric case, with the
$a$-function presented in Section 4 (at least up to two loops).
The reduction of the non-supersymmetric theory presented in Section 2 to the
supersymmetric case (with $n_{\psi}=n_V+n_C$, $n_{\varphi}=2n_C$)
may be accomplished by writing
\begin{align}
\vphi_a&\rightarrow \begin{pmatrix}
\phi_i\\  {\bar \phi}^i
\end{pmatrix}, \quad {\bar \phi}{}^i= ({\phi_i})^* \, ,\quad
\psi_i\rightarrow \begin{pmatrix}
\psi_i\\ \lambda_A
\end{pmatrix},\quad i=1\ldots n_C \, ,
\end{align}
and with $y_a \vphi_a = y^i \phi_i + {\bar y}_i \phibar^i $, 
\begin{align}
y^i&\rightarrow \begin{pmatrix} Y^{ijk}& 0&0&0\\
0&0&0&0\\0&0&0& \sqrt2g(R_B)_j{}^i\\0&0&\sqrt2g(R_A^T)^i{}_k&0
\end{pmatrix},\nn
{\bar y}_i&\rightarrow \begin{pmatrix} 0& \sqrt2g(R_B^T)^j{}_i&0&0\\
\sqrt2g(R_A)_i{}^k&0&0&0\\0&0&\bY_{ijk}&0\\0&0&0&0
\end{pmatrix},
\label{ysup}
\end{align}
where $\lambda$ is the gaugino field. ${\hat {\bar y}}{}_i$ and $\yhat^i$ may be obtained
from ${\bar y}_i$ and $y^i$ by interchanging the upper left and lower right
$2\times2$ blocks of the $4\times4$ matrices. We also have
\be
t_A^{\vphi}\rightarrow\begin{pmatrix}R_A&0\\0&-R_A^T\end{pmatrix},
\quad
t_A^{\psi}\rightarrow\begin{pmatrix}R_A&0\\
0&R_A^{\rm ad}
\end{pmatrix},\quad (R_A^{\rm ad})_{BC}=-if_{ABC} \, ,
\label{tsup}
\ee
and consequently, from Eq.~(\ref{Rdefs}),
\be
R^{\vphi}\rightarrow 2T_R, \quad R^{\psi}\rightarrow T_R+C_G \,  .
\ee
The scalar potential is now given by
\be
V=\tfrac14 \, \bY_{ijm}Y^{klm}{\bar \phi}^i\phibar^j\phi_k\phi_l-g^2\tfrac12
(\phibar R_A \phi)(\phibar R_A \phi) \, .
\label{Vsup}
\ee

In making the reduction from the general theory to the supersymmetric case,
we must start from
two-loop $\beta$-functions corresponding to DRED, since the RG
functions used in Section 3 were evaluated using this scheme; as we mentioned
earlier, the DRED and NSVZ schemes coincide up to the two-loop order we are
considering in this Section.
We use the results given in
Ref.~\cite{JackPL}, which may be obtained from the DREG results by a
coupling redefinition as in Eq.~(\ref{redefgen}) given by
\begin{align}
\mu_4=-\tfrac12,\quad \mu_5=1,\quad \nu_1=\tfrac16,
\label{transf}
\end{align}
with all other coefficients set to zero.\footnote{In general the DRED
$\beta$-functions obtained using \eqref{transf} do not correspond to
a diagrammatic calculational scheme. There is an alternative implementation
of DRED based on a calculational scheme, but involving the use of additional
``evanescent'' couplings. It is this scheme we referred to specifically as
DRED in Ref.~\cite{JackPL}. The two versions of DRED agree in the case of
supersymmetry, which is our focus of interest here; but see
Refs.~\cite{thooft, jjr} for further discussion of the general case.}
\begin{table}[h]
        \setlength{\extrarowheight}{1cm}
        \setlength{\tabcolsep}{24pt}
        \hspace*{-4cm}
        \centering
        \resizebox{6.5cm}{!}{
                \begin{tabular*}{20cm}{cccc}
                        \begin{picture}(162,162) (191,-159)
                        \SetWidth{1.0}
                        \SetColor{Black}
                        \Arc(272,-78)(80,143,503)
                        \Arc(203.807,-43.242)(55.462,-99.136,45.035)
                        \Arc(331.603,-46.277)(50.795,125.647,290.029)
                        \Arc[clock](269.702,-148.829)(53.038,167.113,9.583)
                        \end{picture}
                        &
                        \begin{picture}(162,162) (191,-159)
                        \SetWidth{1.0}
                        \SetColor{Black}
                        \Arc(272,-78)(80,143,503)
                        \Line(272,1)(272,-157)
                        \Arc[clock](182.762,-79.296)(66.239,59.881,-58.876)
                        \Arc(363.91,-81.295)(67.949,120.915,235.066)
                        \end{picture}
                        &
                        \begin{picture}(162,163) (191,-149)
                        \SetWidth{1.0}
                        \SetColor{Black}
                        \Arc(272,-67)(80,143,503)
                        \CBox(282.817,-157.817)(261.183,-136.183){Black}{Black}
                        \Arc(203.807,-32.242)(55.462,-99.136,45.035)
                        \Arc(331.603,-35.277)(50.795,125.647,290.029)
                        \end{picture}
                        &
                        \begin{picture}(173,162) (191,-159)
                        \SetWidth{1.0}
                        \SetColor{Black}
                        \Arc(272,-78)(80,143,503)
                        \CBox(362.817,-90.817)(341.183,-69.183){Black}{Black}
                        \Arc[clock](182.762,-79.296)(66.239,59.881,-58.876)
                        \Arc(363.91,-81.295)(67.949,120.915,235.066)
                        \end{picture}
                        \\
                        {\Huge $G^{S}_{1}$}
                        &
                        {\Huge $G^{S}_{2}$}
                        &
                        {\Huge $G^{S}_{3}$}
                        &
                        {\Huge $G^{S}_{4}$}
                        \\
                        \vspace*{1cm}
                        \\
                        \begin{picture}(167,162) (191,-159)
                        \SetWidth{1.0}
                        \SetColor{Black}
                        \Arc(272,-78)(80,143,503)
                        \CBox(355.817,-120.817)(334.183,-99.183){Black}{Black}
                        \Arc[clock](182.762,-79.296)(66.239,59.881,-58.876)
                        \CBox(356.817,-62.817)(335.183,-41.183){Black}{Black}
                        \end{picture}
                        &
                        \begin{picture}(183,162) (181,-159)
                        \SetWidth{1.0}
                        \SetColor{Black}
                        \Arc(272,-78)(80,143,503)
                        \CBox(362.817,-90.817)(341.183,-69.183){Black}{Black}
                        \Line(272,1)(272,-157)
                        \CBox(203.817,-90.817)(182.183,-69.183){Black}{Black}
                        \end{picture}
                        &
                        \begin{picture}(173,162) (191,-159)
                        \SetWidth{1.0}
                        \SetColor{Black}
                        \Arc(272,-78)(80,143,503)
                        \CBox(362.817,-90.817)(341.183,-69.183){Black}{Black}
                        \Arc[clock](182.762,-79.296)(66.239,59.881,-58.876)
                        \end{picture}
                        &
                        \begin{picture}(162,163) (191,-158)
                        \SetWidth{1.0}
                        \SetColor{Black}
                        \Arc(272,-77)(80,143,503)
                        \Arc[clock](220.79,-157.072)(111.636,102.304,66.11)
                        \Arc[clock](261.627,-112.294)(53.728,66.559,34.322)
                        \Arc[clock](285.384,-122.802)(40.039,40.123,-23.863)
                        \Arc(233.561,-34.52)(94.719,-105.029,9.431)
                        \Line(272,4)(272,-114)
                        \Line(272,-157)(272,-129)
                        \end{picture}
                        \\
                        {\Huge $G^{S}_{5}$}
                        &
                        {\Huge $G^{S}_{6}$}
                        &
                        {\Huge $G^{S}_{7}$}
                        &
                        {\Huge $G^{S}_{8}$}

                \end{tabular*}
        }
        \caption{Contributions to $A^{(4)}$ in the supersymmetric case}
        \label{fig8}
\end{table}
We list here the values of the coefficients in Eq.~(\ref{bcoeffs})
which change under this redefinition (as may be easily checked
using Eqs.~(\ref{redefres}), (\ref{bcoeffs})):
\begin{align}
c^{\rm DR}_6=2\quad
\quad c_{7}^{\rm DR}=-1,\quad
\quad c_{13}^{\rm DR}&=-\tfrac54,\quad
\quad c_{14}^{\rm DR}=\tfrac{1}{4},\quad
\quad c^{\rm DR}_{16}=\tfrac72,\nn
\quad c_{19}^{\rm DR}=7,&\quad
c_{20}^{\rm DR}=-\tfrac{1}{12}\left(138C_G-12R^{\psi}-9R^{\vphi}\right),\nn
c_{26}^{\rm DR}=\tfrac72,&\quad
c_{28}^{\rm DR}=\tfrac{1}{12}(59C_G+4R^{\psi}+R^{\vphi}),
\label{DRcoeffs}
\end{align}

The DRED $\beta$-function leads to the following
alterations in the coefficients in Eq.~(\ref{Afour});
the others are the same as in Eq.~(\ref{Afourres}).
\begin{align}
A^{\rm DR}_{13}=-\tfrac{5}{24},\quad
A^{\rm DR}_{14}=-\tfrac{1}{12},\quad
A^{\rm DR}_{15}&=\tfrac{7}{12},\quad
A^{\rm DR}_{16}=-\tfrac13,\quad
A^{\rm DR}_{17}=\tfrac{1}{6},\nn
A^{\rm DR}_{19}=\tfrac{1}{6},\quad
A^{\rm DR}_{20}=\tfrac{5}{48},\quad
A^{\rm DR}_{22}&=\tfrac14,\quad
A^{\rm DR}_{23}=\tfrac12,\quad
A^{\rm DR}_{25}=-\tfrac52,\nn
A^{\rm DR}_{26}&=-\tfrac{1}{72}[138C_G-12R^{\psi}-9R^{\vphi}]-t_1\beta_0,\nn
A^{\rm DR}_{27}&=\tfrac{1}{144}[59C_G+4R^{\psi}+R^{\vphi}]-t_2\beta_0,
\label{AfourDRED}
\end{align}
These changes are a consequence of making the transformations
Eq.~(\ref{transf})
and also
\be
\delta g=-\frac{g^3}{72n_V} \, \tr[y_a\yhat_a\Chat_R^{\psi}] \, ,
\label{transfa}
\ee
upon $A^{(3)}$ and $A^{(2)}$ respectively in Eq.~(\ref{Alow}). Presumably the
transformation in Eq.~(\ref{transfa}) represents a part of the two-loop
transformation from DREG to DRED (namely the Yukawa dependent
contribution to the transformation of $g$). To the best of our knowledge this
has not been computed in full, though results have been given for the pure
gauge case in Ref.~\cite{lud}.

Inserting Eqs.~(\ref{ysup}), (\ref{tsup}), (\ref{Vsup}) into the expressions
for the various contributions to $\Atil^{(4)}$, depicted in
Table~(\ref{fig8}), we find
\begin{align}
G^A_{1} &\rightarrow 2(G^S_1+9G^S_3-6G^S_4+6G^S_5-12G^S_6
+3(C_G-2T_R)G^S_7+4G^S_8\ldots)\nn
G^A_{2} &\rightarrow 6(G^S_2-2G^S_3+4G^S_4-6G^S_5-(C_G-2T_R)G^S_7+\ldots)\nn
G^A_{3} &\rightarrow 5G^S_3+16G^S_5-16G^S_6-4C_GG^S_7+2G^S_8+\ldots\nn
G^A_{4} &\rightarrow 2(9G^S_5-6G^S_6-3C_GG^S_7+G^S_8+\ldots),\quad
G^A_{5} \rightarrow 2(G^S_1+6G^S_3+12G^S_5+\ldots)\nn
G^A_{6} &\rightarrow 2(G^S_2+4G^S_4+4G^S_6+8T_RG^S_7+\ldots),\nn
G^A_{7} &\rightarrow 2G^S_2+4G^S_3+12G^S_4+24G^S_5+16G^S_6
+8T_RG^S_7+\ldots,\nn
G^A_{8} &\rightarrow 2G^S_1+24G^S_3+96G^S_5+\ldots,\nn
G^A_{9} &\rightarrow 2(-G^S_3-2G^S_4-2G^S_5-4G^S_6-2T_RG^S_7+\ldots),\quad
G^A_{10} \rightarrow
2(G^S_3+4G^S_5+\ldots)\nn
G^A_{11} &\rightarrow 4(-2G^S_3-8G^S_5+\ldots),\quad
G^A_{12} \rightarrow 6(G^S_4-2G^S_5+\ldots)\nn
G^A_{13} &\rightarrow 2(G^S_3+4G^S_5+\ldots),\quad
G^A_{14} \rightarrow 2(G^S_4+2G^S_6+2C_GG^S_7+\ldots)\nn
G^A_{15} &\rightarrow 2(G^S_4+2G^S_5+2G^S_6+\ldots)\quad
G^A_{16} \rightarrow 2(-G^S_5-2G^S_6-C_GG^S_7+\ldots),\quad
G^A_{17} \rightarrow- 8G^S_5+\ldots\nn
G^A_{18} &\rightarrow \tfrac12G^S_3+2C_GG^S_7+\ldots,\quad
G^A_{19} \rightarrow 2G^S_4+4G^S_5+8G^S_6+4C_GG^S_7+\ldots,\nn
G^A_{20} &\rightarrow 2G^S_3+16G^S_5 \ldots,\quad
G^A_{21} \rightarrow 3G^S_5-2G^S_6-\tfrac12C_GG^S_7+\ldots,\quad
G^A_{22} \rightarrow 2G^S_5+\ldots\nn
G^A_{23} &\rightarrow 2G^S_6+\ldots,\quad G^A_{24} \rightarrow
2G^S_5+\ldots,\quad
G^A_{25} \rightarrow 2G^S_6+\ldots\nn
G^A_{26} &\rightarrow 2G^S_7+\ldots,\quad G^A_{27} \rightarrow 2G^S_7+\ldots,
\nn
G^A_{28} &\rightarrow
\tfrac32G^S_1+3G^S_2-12G^S_3-24G^S_4+24G^S_5+48G^S_6
+\ldots \, , &
\end{align}
where again we do not display the purely gauge-coupling dependent
terms.

It is then straightforward to show, using the DRED values
of the coefficients from Eqs.~(\ref{Afourres}),
(\ref{AfourDRED}) that $\Atil^{(4)}$ reduces as
\be
\Atil^{(4)}\rightarrow
\tfrac13\tr[(\gamma^{(1)})^3] + \left(\alpha
- \tfrac{1}{36}\right)\beta^{(1)}_Y \cirk\beta^{(1)}_{\bY}
 + \left[-\tfrac12+8(t_1+t_2)\right]\beta^{(1)}_gg\tr[PC_R] +\ldots\, ,
\label{ourexact}
\ee
where the ellipsis represents pure gauge terms which were not captured by the methods used in Section 2.  
This expression can readily be shown, with the aid of Eqs.~(\ref{Eea}),
(\ref{betag}), to be equivalent at this order to
Eq.~(\ref{Aexact}). The explicit form at this order was already given in
Ref.~\cite{SusyA}; with our notation and conventions, this corresponds
to $\Lambda^{(1)}= Y $ as in Eq.~(\ref{lamonea}) and with
\be
\beta_Y\cirk H \cirk \beta_{\bY}=\alpha \, \beta_Y\cirk\beta_{\bY} \, ,
\ee
and
\be
\lambda=g^2+\tlambda g^4Q+\frac{\tlambda_1}{n_V}g^4\tr[PC_R]+\ldots \, ,
\ee
where we have picked out the terms which can contribute to $Q\tr[PC_R]$.
We find using Eqs.~(\ref{betag}), (\ref{lamone}), (\ref{Lamfour}),
(\ref{gamvals}), (\ref{ourexact}) that we require
\be
t_1+t_2=\frac{29}{16} \, .
\label{tfinal}
\ee
These coefficients correspond to a three-loop calculation
(see Eq.~(\ref{gradfour})) and, in view of Eq.~(\ref{Lamfour}), depend
on the value of $\gamma_{17}$, which has a different value for the
NSVZ scheme than for DRED. It is beyond the scope of this article to
consider how Eq.~(\ref{tfinal}) would be modified within DRED or indeed
within DREG. Since our whole approach is predicated on
the NSVZ scheme, it would probably be naive to assume that the DRED form of
Eq.~(\ref{tfinal}) would be obtained simply by using the DRED result for
$\gamma_{17}$.

Eq.~(\ref{ourexact}) extends the result of Eq.~(7.30) in Ref.\cite{OsbJacnew}
(with $a=3\alpha-\tfrac{1}{12}$) to the gauge case--once again, modulo pure
gauge terms. We see
again the ambiguity in the form of $\Atil$ expressed in general by
Eq.~(\ref{ambig}).

Of course this check is guaranteed to work but nevertheless given the
indirect manner in which we have obtained $\Atil$ and the possibility
of subtleties regarding scheme dependence, it is satisfying to ``close the
loop'' in this fashion.

Finally, we remark that although
the form for $\Atil$ presented in Eq.~(\ref{ourexact}) is
appealingly simple (arguably even more so than Eq.~(\ref{Aexact})),
the obvious extension to higher loops does not appear to be viable.

\section{Conclusions}

In this article we have extended the results of Ref.~\cite{OsbJacnew} to
the case of general gauge theories. In the non-supersymmetric case we have
constructed the terms in the four-loop $a$-function
containing Yukawa or scalar contributions, using the two-loop Yukawa
$\beta$-function and one-loop scalar $\beta$-function. Our main result here
is Eq.~(\ref{Afour}) with Eq.~(\ref{Afourres}). This enabled a
comparison with similar terms in the three-loop gauge $\beta$-function.
In general, as a consequence of the properties of the coupling-constant metric,
 one can
obtain information on the $(n+1)$-loop gauge $\beta$-function from the
$n$- (and lower-) loop Yukawa $\beta$-function and the $(n-1)$ (and lower)
loop scalar $\beta$-function. This is reminiscent of the way in which the
$(n+1)$-loop gauge $\beta$-function is determined by the lower order anomalous
dimensions in a supersymmetric theory, via the NSVZ formula.

In the supersymmetric case we have given a general sufficient condition for
the exact $a$-function of Refs.\cite{AnselmiAM, BarnesJJ, KutasovXU},
given in Eq.~(\ref{Aexact}),
to be valid, and shown that it is satisfied by the three-loop anomalous
dimension. This condition is displayed in Eq.~(\ref{detL}) and is our main
result for the supersymmetric case.

One feature of interest is that Eq.~(\ref{detL}) imposes extra conditions
on the anomalous dimension beyond the mere requirements of integrability
from Eq.~(\ref{grad}); but which are nevertheless satisfied by the explicit 
results as computed. Indeed we remark here (without giving further details
since it is beyond our remit in this article on the gauged case) that
we have observed similar features in the Wess-Zumino model at four loops,
using the results of Ref.\cite{FerreiraRC}.

These properties certainly hint that there might be some underlying reason
why Eq.~(\ref{detL}) {\it must} be satisfied; it would be interesting to
explore
this further. If this were indeed the case, one could imagine exploiting
Eq.~(\ref{detL}) to expedite higher-order calculations of the anomalous
dimension
such as the full gauged case at four loops; possibly combined with
additional information such as the necessary vanishing of $\gamma$ in the
$N=2$ case. Unfortunately, a preliminary check indicates that these constraints
are far from sufficient to determine $\gamma$ completely, even at three
loops; and therefore a considerable quantity of perturbative calculation
would still be unavoidable.

Finally, in Ref.\cite{OsbJacnew} we explored in some detail the freedoms
to redefine the various quantities we have considered, and it would be
interesting to extend these discussions to the current gauged case. In
particular it would be useful to extend Eq.~(\ref{detL}), which in its current
form is predicated upon the NSVZ renormalisation scheme, to a form valid
for {\it any} scheme.

\section{Acknowledgements}
We are very grateful to Tim Jones and Hugh Osborn for useful conversations
and for a careful reading of the manuscript.
CP thanks the STFC for financial support.

\end{document}